\documentclass[prd,preprintnumbers,twocolumn,nofootinbib]{revtex4}
\usepackage{epsfig}
\usepackage{amsmath}
\usepackage{hyperref}
\usepackage{multirow}
\usepackage[utf8]{inputenc}
\usepackage{tikz}
\usetikzlibrary{decorations.pathmorphing}
\usetikzlibrary{decorations.markings}
\usepackage{tkz-euclide}
\usetikzlibrary{shapes,arrows,positioning}
\usepackage[normalem]{ulem}

\begin{document}

\preprint{Cavendish--HEP--25/05}
\renewcommand{\thefigure}{\arabic{figure}}

\title{Untangling the Integration-By-Parts Equations}

\author{Junhan W. Liu}
\email{jl2196@hep.phy.cam.ac.uk}
\affiliation{{\small Cavendish Laboratory, University of Cambridge, Cambridge CB3 0HE, UK}}
\author{Alexander Mitov}
\email{adm74@cam.ac.uk}
\affiliation{{\small Cavendish Laboratory, University of Cambridge, Cambridge CB3 0HE, UK}}

\date{\today}

\begin{abstract}
In this work, we present an algorithm which converts the Integration-by-Parts (IBP) equations into single-variable recurrence relations. We refer to this procedure as {\it diagonalization}. Diagonalized IBP equations are indispensable for reducing loop integrals with extremely high numerator powers to master integrals, and for solving IBP identities in closed analytic form. A prime example is provided by multivariate Mellin representations of loop amplitudes and cross sections. The extension of these methods to other multivariate recurrence relations is also discussed. As a by-product of our diagonalization procedure, we show how the IBP equations can be cast into an efficient, fully triangular form that is well suited for computer implementation. Several complicated topologies have been computed.
\end{abstract}
\maketitle

\section{Introduction\label{sec:intro}}

The Integration-by-Parts Identities (IBP) for Feynman integrals \cite{Tkachov:1981wb,Chetyrkin:1981qh} have been at the forefront of multi-loop amplitude calculations for over four decades. Countless results have been derived with their help, see for e.g. the reviews \cite{Smirnov:2004ym,Weinzierl:2022eaz}. It is therefore astonishing that such a centerpiece of amplitude calculations still lacks an analytic solution and the general results known about the IBPs are very few \cite{Lee:2008tj,Lee:2013hzt}. 

The standard approach for solving the IBPs is to use Gauss elimination \cite{Laporta:2000dsw}, which has been implemented in several computer programs \cite{vonManteuffel:2012np,Lee:2012cn,Wu:2023upw,Maierhofer:2017gsa,Klappert:2020nbg,Lange:2025fba,Smirnov:2025prc}. Further refinements have been proposed, among them Gr$\ddot{\rm o}$bner bases \cite{Smirnov:2005ky,Smirnov:2006tz,Barakat:2022qlc}, Syzygies \cite{Gluza:2010ws,Schabinger:2011dz,Chen:2015lyz,Larsen:2015ped,Bohm:2017qme}, the block-triangular equations \cite{Guan:2019bcx,Guan:2024byi}, the Baikov representation \cite{Baikov:1996rk,Baikov:1996iu}, Intersection Theory \cite{Mastrolia:2018uzb} and most recently Symbolic Reduction Rules \cite{Feng:2025leo,Feng:2026imq,Smith:2025xes}. Methods that solve the IBPs numerically, especially in terms of finite fields and related approaches \cite{vonManteuffel:2014ixa,Peraro:2016wsq,Peraro:2019svx,Klappert:2019emp,Klappert:2020aqs,Chawdhry:2023yyx} have become very popular in the recent past. Very little work on analytic approaches to solving IBPs exists, mainly related to single variable problems \cite{Laporta:2000dsw,Mitov:2005ps,Mitov:2006wy,Kosower:2018obg}.

As implied by our discussion, at present no approach for solving the IBP identities in abstract form exists. To illustrate what we mean by {\it abstract} solution (which we will interchangeably call {\it analytic})  let us consider a simple example, the Fibonacci sequence:
\begin{equation}
F(n) = F(n-1)+F(n-2)\,,
\label{eq:Fibonacci}
\end{equation}
which is fully specified by the two boundary conditions $F(0)=0$ and $F(1)=1$. 

What does it mean to solve the Fibonacci sequence eq.~(\ref{eq:Fibonacci})? For a recurrence relation like eq.~(\ref{eq:Fibonacci}) we will distinguish two approaches: {\it abstract} and {\it numeric}. This terminology is a bit of a misnomer because both approaches produce exact results. The numeric approach allows one to derive $F(n)$ for any specific integer value of $n$ by repeatedly applying the recursion eq.~(\ref{eq:Fibonacci}). This approach is well suited for small to moderate values of $n$. 

The analytic approach means that one derives a solution in closed analytic form. The result for the Fibonacci recursion is well known:
\begin{equation}
F(n) = \frac{\varphi^n - (1-\varphi)^n}{\sqrt{5}}\,,
\label{eq:Fibonacci-closed-form}
\end{equation}
with $\varphi=(1+\sqrt{5})/2$ being the golden ratio. The analytic approach becomes indispensable if one wants to find $F(n)$ for very large or non-integer values of $n$.

Eq.~(\ref{eq:Fibonacci-closed-form}) exhibits properties which are quite relevant for our subsequent discussion: the equation has integer coefficients, its solution for any integer-valued $n$ is an integer, yet its closed form solution is not a rational function of $n$, and it even contains an explicit irrational number. 

The IBP identities differ from the simple recurrence eq.~(\ref{eq:Fibonacci}) by the fact they are multivariate, i.e. they involve several indices $n_1,\dots$. A simple, well-known example of such a multivariate recursion is the so-called contiguous relations satisfied by the Gauss hypergeometric function \cite{BatemanHTF}:
\begin{eqnarray}
0 &=& (c-a-b)F(a,b,c) + a(1-z)F(a+1,b,c) - \nonumber\\
&& (c-b)F(a,b-1,c)\,,\label{eq:F21-a}\\
0 &=& (c-a-b)F(a,b,c) + (a-c)F(a-1,b,c) + \nonumber\\
&& b(1-z)F(a,b+1,c)\,,  \label{eq:F21-b}\\
0 &=& (c-a-1)F(a,b,c) + aF(a+1,b,c) - \nonumber\\
&& (c-1)F(a,b,c-1)\,,\label{eq:F21-c}
\end{eqnarray}
where for brevity we denote ${}_2F_1(a,b,c,z) \equiv F(a,b,c)$.

In this paper, we refer to equations of the type of eq.~(\ref{eq:Fibonacci}) as diagonal equations, since they define recurrence relations that shift only a single index. Likewise, we refer to equations of the type of eqs.~(\ref{eq:F21-a},\ref{eq:F21-b},\ref{eq:F21-c}) as non-diagonal equations, since they define recurrence relations that shift multiple indices simultaneously.

Concerning the IBP equations, as already apparent from eqs.~(\ref{eq:F21-a},\ref{eq:F21-b},\ref{eq:F21-c}), their non-diagonal form is perhaps the single largest impediment for their solving. It significantly increases the effort required for solving the system numerically when high numerator powers are present, and prevents the application of conventional techniques for analytically solving recurrence relations.

In this work we develop an algorithm which allows the systematic diagonalization of IBP identities. To the best of our knowledge, this is the first time such a result has been achieved. For reasons we will describe later, we will present two ways of writing an IBP system in a diagonal form. Being completely generic, our diagonalization algorithm can be applied to any system of linear homogeneous recurrence relations that admits a finite basis. For example, in Appendix \ref{app:F21} we apply it to the contiguous relations eqs.~(\ref{eq:F21-a},\ref{eq:F21-b},\ref{eq:F21-c}), and diagonalize them to show they require two boundary conditions. We also recover well-known properties of this classic function solely from its contiguous relations.

This paper is organized as follows: in sec.~\ref{sec:diag} we introduce the main result of this work, namely, three different ways of writing a system of IBP equations. In sec.~\ref{sec:algo} we present our algorithm in full generality. In sec.~\ref{sec:benchmarks} we present a set of benchmarks results achieved with our algorithm and show how it compares with current state-of-the-art algorithms for solving the IBP identities. Our conclusions are presented in sec.~\ref{sec:conclusions}. In appendix~\ref{app:F21} we present some results for the contiguous relations eqs.~(\ref{eq:F21-a},\ref{eq:F21-b},\ref{eq:F21-c}), in appendix~\ref{app:fibonacci} we derive and discuss the Fibonacci recurrence in matrix form, and in appendix~\ref{app:diageq} we specify the diagonal equations for the one-loop box topology.

\section{Diagonalizing the IBP identities}\label{sec:diag}

The diagonalization procedure proposed in this paper is quite involved. To make the discussion more transparent, in this section we present all results using a simple example: the one-loop box topology. This choice keeps the main ideas and features clear while being general enough to exhibit the essential features we have observed even in the most complicated cases we have considered.

The one-loop box topology is depicted in fig.~\ref{fig:box}. It is defined through the following set of four propagators $P_i$
\begin{equation}
 P_1=k^2,\,P_2=(k-p_1)^2,\,P_3=(k-p_1-p_2)^2,\,P_4=(k+p_3)^2\,.
\label{eq:propagators}
\end{equation}
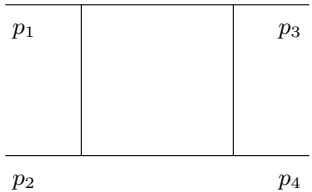
\begin{figure}[t]
\vskip 5mm
\centering
\begin{tikzpicture}[scale=0.5]
\node (p1) at  (-3.5,1.3)  {$p_1$};
\node (p2) at  (-3.5,-2.7) {$p_2$};
\node (p3) at  (3.5,1.3)   {$p_3$};
\node (p4) at  (3.5,-2.7)  {$p_4$};
\draw  (-4,2) to (4,2) {};
\draw  (-4,-2) to (4,-2) {};
\draw  (-2,-2) to (-2,2) {};
\draw  (2,-2) to (2,2) {};
\end{tikzpicture}
\caption{The one-loop box topology.}
\label{fig:box}
\end{figure}

All Feynman integrals for this topology can be parametrized as 
\begin{equation}
I_{\nu_1, \nu_2, \nu_3, \nu_4} = \int d^dk \frac{1}{P_1^{\nu_1}P_2^{\nu_2}P_3^{\nu_3}P_4^{\nu_4}}\,.
\label{eq:I}
\end{equation}
They depend on the space-time dimension $d$ and the following kinematic variables:
\begin{eqnarray}
&& 2p_1 \cdot p_2 = s,\; 2p_1 \cdot p_3 = st,\; 2p_2 \cdot p_3 = -s(t +1)\,,\\
&& p_1^2 = 0,\; p_2^2 = 0,\; p_3^2 = 0\,,
\label{eq:kin}
\end{eqnarray}
where $t$ is dimensionless. In the following we set $s=1$ for brevity. The dependence on $s$ for each integral can be easily reconstructed on dimensional grounds. 

There are four IBP identities for this topology:
\begin{eqnarray}
0 &=& -\nu_4 I_{\nu_1 - 1, \nu_2, \nu_3, \nu_4 + 1} - \nu_3 I_{\nu_1 - 1, \nu_2, \nu_3 + 1, \nu_4}\nonumber\\
&& - \nu_2 I_{\nu_1 - 1, \nu_2 + 1, \nu_3, \nu_4} + \nu_3\, I_{\nu_1, \nu_2, \nu_3 + 1, \nu_4} \nonumber\\
&& + (d - 2\nu_1 - \nu_2 - \nu_3 - \nu_4) I_{\nu_1, \nu_2, \nu_3, \nu_4}\,,\label{eq:IBP-1-loop-box-1}\\
0 &=& -\nu_4 I_{\nu_1 - 1, \nu_2, \nu_3, \nu_4 + 1} - \nu_3 I_{\nu_1 - 1, \nu_2, \nu_3 + 1, \nu_4}\nonumber\\
&& - \nu_2 I_{\nu_1 - 1, \nu_2 + 1, \nu_3, \nu_4} + \nu_4 I_{\nu_1, \nu_2 - 1, \nu_3, \nu_4 + 1}\nonumber\\
&& + \nu_3 I_{\nu_1, \nu_2 - 1, \nu_3 + 1, \nu_4} - \nu_4 t\, I_{\nu_1, \nu_2, \nu_3, \nu_4 + 1} \nonumber\\
&& + \nu_3\, I_{\nu_1, \nu_2, \nu_3 + 1, \nu_4} + \nu_1 I_{\nu_1 + 1, \nu_2 - 1, \nu_3, \nu_4}\nonumber\\
&& + (\nu_2 - \nu_1) I_{\nu_1, \nu_2, \nu_3, \nu_4}\,,\label{eq:IBP-1-loop-box-2}\\
0 &=& -\nu_4 I_{\nu_1, \nu_2 - 1, \nu_3, \nu_4 + 1} - \nu_3 I_{\nu_1, \nu_2 - 1, \nu_3 + 1, \nu_4}\nonumber\\
&& + \nu_4 I_{\nu_1, \nu_2, \nu_3 - 1, \nu_4 + 1} + \nu_4 t\, I_{\nu_1, \nu_2, \nu_3, \nu_4 + 1}\nonumber\\
&& + \nu_2 I_{\nu_1, \nu_2 + 1, \nu_3 - 1, \nu_4} - \nu_1 I_{\nu_1 + 1, \nu_2 - 1, \nu_3, \nu_4}\nonumber\\
&& + \nu_1 I_{\nu_1 + 1, \nu_2, \nu_3 - 1, \nu_4} - \nu_1\, I_{\nu_1 + 1, \nu_2, \nu_3, \nu_4}\nonumber\\
&& + (\nu_3 - \nu_2) I_{\nu_1, \nu_2, \nu_3, \nu_4}\,,\label{eq:IBP-1-loop-box-3}\\
0&=& \nu_4 I_{\nu_1 - 1, \nu_2, \nu_3, \nu_4 + 1} + \nu_3 I_{\nu_1 - 1, \nu_2, \nu_3 + 1, \nu_4}\nonumber\\
&& + \nu_2 I_{\nu_1 - 1, \nu_2 + 1, \nu_3, \nu_4}  - \nu_3 I_{\nu_1, \nu_2, \nu_3 + 1, \nu_4 - 1}\nonumber\\
&& - \nu_3\, I_{\nu_1, \nu_2, \nu_3 + 1, \nu_4} - \nu_2 I_{\nu_1, \nu_2 + 1, \nu_3, \nu_4 - 1}\nonumber\\
&& + \nu_2 t\, I_{\nu_1, \nu_2 + 1, \nu_3, \nu_4} - \nu_1 I_{\nu_1 + 1, \nu_2, \nu_3, \nu_4 - 1}\nonumber\\
&& + (\nu_1 - \nu_4) I_{\nu_1, \nu_2, \nu_3, \nu_4}\,.
\label{eq:IBP-1-loop-box-4}
\end{eqnarray}

As one can see, even in this very simple topology the set of IBP identities is quite involved and all equations are highly non-diagonal.

\subsection{Equations in diagonal form}\label{sec:strict-diag}

By applying the algorithm described in sec.~\ref{sec:algo}, we derive a set of diagonal recurrence relations. As their name suggests, these relations shift one index at a time, while keeping all other indices fixed. Starting with the index $\nu_1$ one arrives at the following third order diagonal recurrence relation:
\begin{equation}
I_{\nu_1-3, \nu_2, \nu_3, \nu_4} = \sum_{k=0}^2 r_k(\nu_1, \nu_2, \nu_3, \nu_4)I_{\nu_1-k, \nu_2, \nu_3, \nu_4}\,,
\label{eq:diag-abstract}
\end{equation}
where $r_k$ are rational functions. Their explicit expressions can be found in appendix~\ref{app:diageq}. Similar relations can be derived for the other indices $\nu_{2,3,4}$.

A central feature of eq.~(\ref{eq:diag-abstract}) is its order. The order of the recurrence relation, three, equals the number of master integrals in the whole topology. The fact that the order of the equations equals the number of masters is to be expected, since for abstract values of the indices -- as is the case for eq.~(\ref{eq:diag-abstract}) -- the equations connect every integral in the full topology. If one tries to solve this diagonal system ``numerically" i.e. for specific integer values of the indices $\nu_i$, the order of the equations will remain the same.

In this section we choose to work with the following basis of masters: $I_{0,1,0,1}$, $I_{1,0,1,0}$ and $I_{1,1,1,1}$.

For a typical calculation of gauge theory amplitudes, one encounters indices in the range $\nu_i \leq 1$. ``Squared" propagators with $\nu_i=2$ appear very rarely and in the following we will ignore $\nu_i=2$ . This is done only for simplicity of presentation; there are various ways one can deal with such propagators, if they are present. The approaches developed in this work do not introduce squared propagators at the intermediate stages, see sec.~\ref{sec:algo-I} for details.

Applying first eq.~(\ref{eq:diag-abstract}) and then the other three diagonal equations, one can map any integral to integrals with $\nu_i=-1,0,1$. Clearly this is a much larger set of integrals which could contain as many as $3^4=81$ integrals. In practice, additional constraints will appear when the diagonal equations are calculated for specific integer values for some of their indices. This will reduce the number of unsolved integrals significantly, although it will still be larger than the expected basis of masters consisting of just three integrals. All but three of the unsolved integrals can be mapped to the three master integrals by performing a new IBP solution that spans the range of indices $\nu_i=-1,0,1$. We note that such a situation has been encountered previously \cite{Mitov:2005ps,Mitov:2006wy}. In sec.~\ref{sec:matrix-diag} we will present a modification of this method, which we call {\it matrix-diagonal}, which does not produce unsolved dependent integrals.

Before concluding this subsection, we would like to summarize the two main advantages of the diagonal approach. First, it allows one to solve the IBP identities for cases where the indices $\nu_i$ are not necessarily integers. Second, it allows one to reduce integrals $I_{\nu_1, \nu_2, \nu_3, \nu_4}$ with very high negative integer values of the indices $\nu_i$, since in such case existing approaches would quickly become impractical.

The diagonal approach also has some disadvantages. For example, it does not map all integrals to master integrals. As a result, an additional - albeit simple - reduction must be performed to reduce the remaining redundant boundary conditions to master integrals. It also may lead to recurrence relations of very high order for topologies with many master integrals. This can be impractical since it may require one to map the topology to integrals with very high numerator powers. As we will see in subsection~\ref{sec:matrix-diag}, the diagonal approach can be recast in a slightly different form which does not have any of its drawbacks.

\subsection{Sector structure and reduction of the order of diagonal recurrences}\label{sec:reduce-order}

In practice, one often encounters situations where integrals have some of their indices fixed. For example, all integrals in the one-loop box topology belong to one of the two bottom groups of sectors $\overline{(0, 1, 0, 1)}$ and $\overline{(1, 0, 1, 0)}$. Here we have introduced the concise notation:
\begin{equation}
\overline{(0, 1, 0, 1)} \equiv (0, 1, 0, 1) ~ +~{\rm all~sectors~above~it}\,.
\end{equation}

This implies that the integrals in this topology can be parametrized with at most two abstract indices as $I_{\nu_1, 1, \nu_3, 1}$ and $I_{1, \nu_2, 1, \nu_4}$. One can easily see that each one of these two bottom groups of sectors contains only two masters, not three. This suggests that it may be possible to reduce the integrals $I_{\nu_1, 1, \nu_3, 1}$ and $I_{1, \nu_2, 1, \nu_4}$ using difference equations of second order, not third. In the rest of this subsection we explain how this may be achieved.

As noted in sec.~\ref{sec:strict-diag}, the order of a generic difference equation like eq.~(\ref{eq:diag-abstract}) cannot be reduced by simply setting some abstract indices $\nu_i$ to integer values. This can be understood as follows. Integrals like $I_{\nu_1, 1, \nu_3, 1}$ with $\nu_i\leq 1$ effectively connect to the top sector of the topology, which then maps to all masters in the topology. In other words, abstract equations with abstract indices do not fully respect the sector structure of the IBPs.

We next explain how one can partially re-impose the sector structure by hand and then use that to derive a recurrence of second order. Consider, for example, the integrals $I_{\nu_1, 1, \nu_3, 1}$ with indices $\nu_{1,3}\leq 1$. These integrals span the sectors $\overline{(0, 1, 0, 1)}$. As mentioned above, there are two masters in $\overline{(0, 1, 0, 1)}$, $I_{0,1,0,1}$ and $I_{1,1,1,1}$, which are relevant only to integrals belonging to this group of sectors. In turn, this implies that masters which belong to sectors outside of $\overline{(0, 1, 0, 1)}$ will be irrelevant for the projection of the integrals $I_{\nu_1, 1, \nu_3, 1}$ to $I_{0,1,0,1}$ and $I_{1,1,1,1}$. In this example there is one such master: $I_{1,0,1,0}$. Following ref.~\cite{Chawdhry:2018awn} we set the master $I_{1,0,1,0}$ to zero. This effectively sets the whole sector $(1, 0, 1, 0)$ to zero, together with its higher sectors which are not located above the sector $(0, 1, 0, 1)$. 

We would like to emphasize that while any one master can be set to zero, see ref.~\cite{Chawdhry:2018awn} for details, this may not always be beneficial in the context of equations with abstract indices. The reason is that not all masters can be identified explicitly in the course of the derivation of the diagonal equations. What one {\it can} identify, however, are integrals belonging to sectors outside the sector being considered. If we take as an example the integrals $I_{\nu_1, 1, \nu_3, 1}$, one can identify -- and therefore set to zero -- all integrals with vanishing second or fourth index. Any such integral, even if it has abstract $\nu_1$ and/or $\nu_3$, necessarily belongs to a sector which vanishes if the master $I_{1,0,1,0}$ is set to zero. 

The key observation here is that the sector structure that can be utilized is defined solely by the indices which are explicitly set to positive numeric values, like the second and fourth index in the integral  $I_{\nu_1, 1, \nu_3, 1}$, not by the indices which are abstract. 

The above observations allow one to derive a recurrence relation for $I_{\nu_1, 1, \nu_3, 1}$ which is of second order, not third. Repeating this procedure also for the integrals $I_{1, \nu_2, 1, \nu_4}$ in $\overline{(1, 0, 1, 0)}$, one can map the full set of integrals in the one-loop box topology to a much reduced set consisting of just five integrals. Only two of them are not independent. They can be reduced to masters once additional relations are derived, for example, with a minimal Laporta calculation. The complete set of second order diagonal equations can be found in appendix~\ref{app:diageq}.

One can understand the two cases, when the integral $I_{\nu_1, 1, \nu_3, 1}$ satisfies a second or a third order recurrence, as follows. To derive the projection of an integral like  $I_{\nu_1, 1, \nu_3, 1}$ onto the sectors $\overline{(0, 1, 0, 1)}$ where there are two masters, one has two options. First, one can solve the full topology which in this example will lead to a third order recurrence. This general solution can then be restricted to the two masters in the sectors $\overline{(0, 1, 0, 1)}$. Alternatively, since one is only concerned with the masters in $\overline{(0, 1, 0, 1)}$, one can set to zero all masters (and therefore sectors) that do not belong to $\overline{(0, 1, 0, 1)}$, and then derive a recurrence relation which would only apply to the sectors $\overline{(0, 1, 0, 1)}$. This equation will be of second order. Importantly, the results from both approaches will be the same, when restricted to the masters in the sectors $\overline{(0, 1, 0, 1)}$. We have confirmed this agreement by direct calculation of several non-trivial two loop topologies.

\subsection{Equations in matrix-diagonal form}\label{sec:matrix-diag}

One can write the diagonal equations shown in the previous section in a slightly different form which, amazingly, does not suffer from any of the shortcomings observed in the diagonal approach. As we will see, this way of formulating the diagonal equations leads directly to the master integrals in the problem, and no dependent integral will remain unsolved. This form seems to also be much more beneficial for attempting analytic solutions to the recurrence relations. 

One way to motivate this approach is to recall that a higher order recurrence equation can be replaced by a system of first order equations. Working in the approach where the solution is derived for each bottom sector in the topology (see sec.~\ref{sec:strict-diag} and \ref{sec:reduce-order} for details) we expect that the second order equations can be written in a $2\times 2$ matrix form. Starting with the sectors $\overline{(0,1,0,1)}$, we choose to parametrize their integrals in the following way:
\begin{equation}
\boldsymbol{V}_{\nu_1,\nu_3}=(I_{\nu_1,1,\nu_3,1},I_{\nu_1-1,1,\nu_3,1})^T\,,
\label{eq:matrix-V}
\end{equation}
with $\nu_{1,3}\leq 1$. Other choices for eq.~(\ref{eq:matrix-V}) are possible. We have found out from experience that choosing the integrals inside the vector $\boldsymbol{V}$ such that their indices are separated as little as possible leads to more compact equations. Furthermore, it is convenient if the lowest values of the indices, namely $\nu_{1,3} = 1$, correspond to the master integrals. With this in mind, in this section we use the following basis of masters:
\begin{equation}
\boldsymbol{V}_{1,1}=(I_{1,1,1,1},I_{0,1,1,1})^T\,.
\label{eq:matrix-basis}
\end{equation}

Following the algorithm presented in sec.~\ref{sec:algo}, we derive the following very compact equations:
\begin{equation}
\boldsymbol{V}_{\nu_1-1,\nu_3}=\boldsymbol{W^{(2)}}(\nu_1,\nu_3) \boldsymbol{V}_{\nu_1,\nu_3}\,,
\label{eq:matrix-nu1}
\end{equation}
and
\begin{equation}
\boldsymbol{V}_{1,\nu_3-1}=\boldsymbol{W^{(1)}}(\nu_3) \boldsymbol{V}_{1,\nu_3}\,,
\label{eq:matrix-nu3}
\end{equation}
where the $2\times 2$ matrices $\boldsymbol{W^{(2)}}(\nu_1,\nu_3)$ and $\boldsymbol{W^{(1)}}(\nu_3)$ read:
\begin{eqnarray}
&&\boldsymbol{W^{(2)}}(\nu_1,\nu_3) = \label{eq:W2}\\
&&\begin{pmatrix}
0&1\\ 
\frac{(2 - d + 2 \nu_1)  (\nu_1-1)  t}{2 (2 - d + \nu_1 + \nu_3) (1 - d + \nu_1 + \nu_3)} & \frac{2 - 2 \nu_1 + 2 \nu_3 - d t + 2 \nu_1 t + 2 \nu_3 t}{2 (-1 + d - \nu_1 - \nu_3)}
\end{pmatrix}\,,\nonumber
\end{eqnarray}
and
\begin{equation}
\boldsymbol{W^{(1)}}(\nu_3) = \begin{pmatrix}
\frac{1-\nu_3}{-3+d-\nu_3}&1\\
0&\frac{(2 - d + 2 \nu_3) t}{2 (-2 + d - \nu_3)}
\end{pmatrix}\,.
\label{eq:W1}
\end{equation}

To this end, one first reduces the index $\nu_1$ down to $\nu_1=1$ and then the second index, $\nu_3$, down to $\nu_3=1$. The final expression for any integral $I_{\nu_1,1,\nu_3,1}$ can be obtained as the upper component of:
\begin{equation}
\boldsymbol{V}_{\nu_1,\nu_3} = \left( \prod_{n=1}^{\nu_1+1}\boldsymbol{W^{(2)}}(n,\nu_3)\right)\left(   \prod_{m=1}^{\nu_3+1} \boldsymbol{W^{(1)}}(m)\right) \boldsymbol{V}_{1,1} \,.
\label{eq:matrix-solution}
\end{equation}
All integrals are therefore reduced to $\boldsymbol{V}_{1,1}$, which are the two master integrals defined in eq. (\ref{eq:matrix-basis}), without any integrals remaining unsolved. A completely analogous result can be derived for the integrals $I_{1, \nu_2, 1, \nu_4}$ belonging to this topology's second bottom group of sectors $\overline{(1, 0, 1, 0)}$.

Before closing this section, we would like to make two remarks. First, the equations above are well suited to be extended to non-integer values for the indices $\nu_i$. As is well known, in the case of first order recurrences, the solution to eq.~(\ref{eq:matrix-solution}) can easily be extended to this domain by expressing the product through Euler $\Gamma$ functions. The form of eq.~(\ref{eq:matrix-solution}) clearly suggests that this would be a natural approach also for the case of higher order recurrences by seeking a suitable matrix generalization of the usual $\Gamma$ function. We postpone a more detailed investigation of this topic for a separate publication.

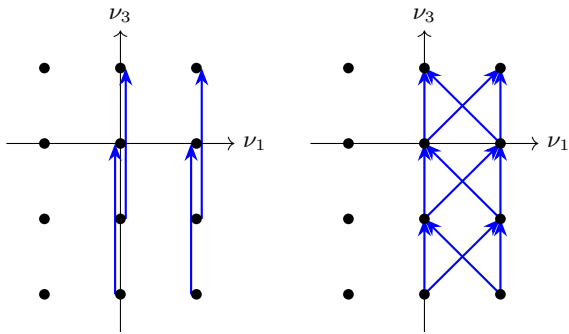
\begin{figure}[t]
\centering
\begin{tikzpicture}
\begin{scope}[shift={(-2,0)}]
    \draw[->] (-1.5,0) -- (1.5,0) node[right] {$\nu_1$};
    \draw[->] (0,-2.5) -- (0,1.5) node[above] {$\nu_3$};
    \draw[->, >=Stealth, blue, thick] (1.07,-1) -- (1.07,1);
    \draw[->, >=Stealth, blue, thick] (0.07,-1) -- (0.07,1);
    \draw[->, >=Stealth, blue, thick] (0.93,-2) -- (0.93,0);
    \draw[->, >=Stealth, blue, thick] (-0.07,-2) -- (-0.07,0);
    \foreach \x/\y in {-1/-2, 0/-2, 1/-2,
                       -1/-1, 0/-1, 1/-1,
                       -1/0,  0/0,  1/0,
                       -1/1, 0/1,  1/1}{
        \fill (\x,\y) circle (2pt);
    }
\end{scope}
\begin{scope}[shift={(2,0)}]
    \draw[->] (-1.5,0) -- (1.5,0) node[right] {$\nu_1$};
    \draw[->] (0,-2.5) -- (0,1.5) node[above] {$\nu_3$};
    \draw[->, >=Stealth, blue, thick] (0,0) -- (1,1);
    \draw[->, >=Stealth, blue, thick] (0,0) -- (0,1);
    \draw[->, >=Stealth, blue, thick] (1,0) -- (1,1);
    \draw[->, >=Stealth, blue, thick] (1,0) -- (0,1);
    \draw[->, >=Stealth, blue, thick] (0,-1) -- (1,0);
    \draw[->, >=Stealth, blue, thick] (0,-1) -- (0,0);
    \draw[->, >=Stealth, blue, thick] (1,-1) -- (1,0);
    \draw[->, >=Stealth, blue, thick] (1,-1) -- (0,0);
    \draw[->, >=Stealth, blue, thick] (0,-2) -- (1,-1);
    \draw[->, >=Stealth, blue, thick] (0,-2) -- (0,-1);
    \draw[->, >=Stealth, blue, thick] (1,-2) -- (1,-1);
    \draw[->, >=Stealth, blue, thick] (1,-2) -- (0,-1);
    \foreach \x/\y in {-1/-2, 0/-2, 1/-2,
                       -1/-1, 0/-1, 1/-1,
                       -1/0,  0/0,  1/0,
                       -1/1, 0/1,  1/1}{
        \fill (\x,\y) circle (2pt);
    }
\end{scope}
\end{tikzpicture}
\caption{An illustration of the ``evolution" of the diagonal (left) and matrix-diagonal (right) equations.}
\label{fig:sweep}
\end{figure}

Second, we would like to clarify why the matrix-diagonal formulation maps every integral directly to the set of master integrals, while its seemingly equivalent diagonal formulation (see sec.~\ref{sec:strict-diag}) leads to a mapping to a wider set of integrals, which requires a further (albeit simple) reduction down to the set of master integrals. This can be understood by looking at fig.~\ref{fig:sweep} which clearly shows the very different ways the two solutions ``evolve" in the corresponding indices. While the diagonal approach modifies only one index at a time, the matrix-diagonal approach mixes integrals at every step, which is why no integral remains unsolved. 

Before we close this section, we would like to discuss a potentially significant feature of eq.~(\ref{eq:matrix-nu1}). This equation looks like a standard matrix transformation and one may ask if it is possible to find an additional transformation matrix, $ \boldsymbol{O}(\nu_1,\nu_3)$, which transforms the basis of integrals to a new basis where the matrix
 \begin{equation}
 \boldsymbol{W'} =  \boldsymbol{O}^{-1}(\nu_1-1,\nu_3) \boldsymbol{W}  \boldsymbol{O}(\nu_1,\nu_3)\,,
 \end{equation}
 is diagonal. If the answer to this question is affirmative, one can easily see that it will have many non-trivial implications. For example, this would imply that the integrals in the new basis satisfy independent first-order recurrence relations. Clearly, such a basis of integrals must be very special and its properties will definitely need to be studied, which is something we would like to return to in a separate publication. 

On the other hand one would not expect a diagonal matrix $\boldsymbol{W'}$ to have elements which are rational functions. The reason for this is that if they were rational, the integrals would be solvable in $\Gamma$ functions only. We know from experience that in general hypergeometric functions appear in such solutions. One can confirm this intuition on a very simple example: the Fibonacci sequence eq.~(\ref{eq:Fibonacci}). Please see appendix \ref{app:fibonacci} for the details.

\subsection{Motivation for equations in triangular form}\label{sec:triangular-motiv}

While the results in the previous subsections show that we have solved the problem posed in this paper, the question of the efficiency of the diagonal equations remains. This is especially the case for the range of indices typically of interest in amplitude calculations. Tests of complicated two-loop topologies show that the diagonal equations become of a very substantial size due to growing rational coefficients appearing in the equations. Although the solution eq.~(\ref{eq:matrix-solution}) of the matrix-diagonal approach is complete and conceptually completely straightforward, the repeated matrix multiplication leads to further growth of rational coefficients. 

While this may be dealt with using other methods -- including finite field reconstruction and related techniques -- it is interesting to ask if one can in principle derive a different approach, one which is optimized for solving typical calculations of amplitudes. Such an approach should have better efficiency than the diagonal approach for small to moderate values of the indices $\nu_i$ and retain as much as possible the simplicity and straightforwardness inherent in the diagonal approach. In particular, we envision its application through existing IBP solving programs. 

In this section we demonstrate that the answer to this question is affirmative. We call such an approach {\it triangular}. In sec.~\ref{sec:benchmarks} we show several benchmark results for this approach and compare it with existing state-of-the-art approaches. However, before we specify the triangular algorithm and how it is applied in practice, it is instructive to first quantify a measure for the efficiency of IBP solving approaches. 

The IBP identities represent an infinite system of linear homogeneous equations. In this work we have come to the realization that it is best to represent the explicit IBP system of equations in the following way:
\begin{equation}
\boldsymbol{\cal F} = \boldsymbol{M}\cdot \boldsymbol{\cal F}\,,
\label{eq:FeqMF}
\end{equation}
where $\boldsymbol{M}$ is an infinite matrix with its elements being rational functions, and $\boldsymbol{\cal F}$ is an infinite ``vector" which contains all integrals in the topology:
\begin{equation}
\boldsymbol{\cal F} = \left({\cal F}_1,\dots,{\cal F}_{n},{\cal F}_{n+1},\dots \right)^T\,.
\label{eq:F}
\end{equation}
The ordering of the integrals inside $\boldsymbol{\cal F}$ is arbitrary. We have found it convenient to first place all relevant master integrals ${\cal F}_1,\dots, {\cal F}_n$, followed by the remaining Feynman integrals ${\cal F}_{n+1},\dots$. Their ordering can be chosen at will, although it is very useful to order them in complexity as, for example, defined by Laporta \cite{Laporta:2000dsw}. In the following we adopt similar convention.

It is most instructive to study the structure of the matrix $\boldsymbol{M}$ for each set of equations. In eq.~(\ref{eq:M}), below, we show the top $10\times 10$ sub-matrices for each of the three types of equations: standard Laporta (left), diagonal (center) and matrix-diagonal (right).

Since our goal is to understand the structure of the matrix $\boldsymbol{M}$, all elements that are neither 0 nor 1 have for brevity been denoted simply as ${\rm x}$, irrespective of their actual values.

There are several crucial observations to be made about the matrices in eq.~(\ref{eq:M}). The matrix $\boldsymbol{M}$ exhibits a very special structure for the diagonal and matrix-diagonal equations. The top left block is the unit $1\times 1$ matrix while the next block along the main diagonal is a strictly lower triangular $9\times 9$ matrix. All matrix elements above the main diagonal vanish. 
\begin{widetext}
\begin{equation}
\begin{pmatrix}
\mathbf{1} & 0 & 0 & 0 & 0 & 0 & 0 & 0 & 0 & 0 \\
{\rm x} & \mathbf{1} & 0 & 0 & {\rm x} & 0 & 0 & {\rm x} & 0 & 0  \\
1 & 0 & \boldsymbol{{\rm x}} & 0 & {\rm x} & 0 & 0 & {\rm x} & 0 & 0  \\
{\rm x} & {\rm x} & 0 & \mathbf{1} & 1 & 0 & 0 & 1 & 0 & 0 \\
{\rm x} & 0 & {\rm x} & 0 & \mathbf{1} & 0 & 1 & 0 & 1 & 0 \\
0 & 0 & 0 & {\rm x} & 0 & \mathbf{1} & 0 & 0 & 0 & {\rm x}  \\
0 & 0 & 0 & {\rm x} & 0 & 0 & \mathbf{1} & {\rm x} & 0 & {\rm x}  \\
0 & 0 & 0 & {\rm x} & {\rm x} & 0 & 0 & \mathbf{1} & 0 & 1  \\
1 & 0 & 0 & {\rm x} & 0 & {\rm x} & 0 & {\rm x} & \mathbf{1} & 0 \\
{\rm x} & 0 & 0 & 1 & 0 & {\rm x} & 0 & 0 & 0 & \mathbf{1}
\end{pmatrix}~\,,~
\begin{pmatrix}
\mathbf{1} & 0 & 0 & 0 & 0 & 0 & 0 & 0 & 0 & 0 \\
{\rm x} & \mathbf{0} & 0 & 0 & 0 & 0 & 0 & 0 & 0 & 0 \\
0 & {\rm x} & \mathbf{0} & 0 & 0 & 0 & 0 & 0 & 0 & 0 \\
0 & 0 & {\rm x} & \mathbf{0} & 0 & 0 & 0 & 0 & 0 & 0 \\
0 & 0 & 0 & {\rm x} & \mathbf{0} & 0 & 0 & 0 & 0 & 0 \\
0 & 0 & 0 & 0 & {\rm x} & \mathbf{0} & 0 & 0 & 0 & 0 \\
0 & 0 & 0 & 0 & 0 & {\rm x} & \mathbf{0} & 0 & 0 & 0 \\
0 & 0 & 0 & 0 & 0 & 0 & {\rm x} & \mathbf{0} & 0 & 0 \\
0 & 0 & 0 & 0 & 0 & 0 & 0 & {\rm x} & \mathbf{0} & 0 \\
0 & 0 & 0 & 0 & 0 & 0 & 0 & {\rm x} & {\rm x} & \mathbf{0}
\end{pmatrix}~\,,~
\begin{pmatrix}
\mathbf{1} & 0 & 0 & 0 & 0 & 0 & 0 & 0 & 0 & 0 \\
1 & \mathbf{0} & 0 & 0 & 0 & 0 & 0 & 0 & 0 & 0 \\
{\rm x} & 0 & \mathbf{0} & 0 & 0 & 0 & 0 & 0 & 0 & 0 \\
0 & {\rm x} & 1 & \mathbf{0} & 0 & 0 & 0 & 0 & 0 & 0 \\
0 & 0 & {\rm x} & 0 & \mathbf{0} & 0 & 0 & 0 & 0 & 0 \\
0 & 0 & 0 & {\rm x} & 1 & \mathbf{0} & 0 & 0 & 0 & 0 \\
0 & 0 & 0 & 0 & {\rm x} & 0 & \mathbf{0} & 0 & 0 & 0 \\
0 & 0 & 0 & 0 & 0 & {\rm x} & 1 & \mathbf{0} & 0 & 0 \\
0 & 0 & 0 & 0 & 0 & 0 & {\rm x} & 0 & \mathbf{0} & 0 \\
0 & 0 & 0 & 0 & 0 & 0 & 0 & {\rm x} & 1 & \mathbf{0}
\end{pmatrix}\,.
\label{eq:M}
\end{equation}
\end{widetext}

A matrix with such a structure has a special property: as it is raised to higher and higher powers, the $9\times 9$ block will eventually become 0 and all non-zero elements will be shifted to the first column of this matrix.

The above observation is very important because in fact it represents a way to solve the IBP identities. By repeatedly applying eq.~(\ref{eq:FeqMF}) one has:
\begin{equation}
\boldsymbol{\cal F} = \boldsymbol{M}\cdot \boldsymbol{\cal F} = \boldsymbol{M}^2\cdot \boldsymbol{\cal F} = \boldsymbol{M}^3\cdot \boldsymbol{\cal F} = \dots \,,
\label{eq:powers}
\end{equation}
i.e. the matrix $\boldsymbol{M}$ can be replaced in the IBP equations by any integer power of itself. Therefore, for matrices that have the above special {\it block-diagonal-lower-triangular} structure, after raising the matrix $\boldsymbol{M}$ to power $m$, the first $m$ elements in $\boldsymbol{\cal F}$ that come after the master integrals will automatically be solved. One could in fact very efficiently reach very large powers $m$ by performing only $k$ matrix multiplications, where $k$ is the smallest integer which satisfies $2^k\ge m$, and {\it use this as yet another way of solving the IBP identities}.

This special block-diagonal-lower-triangular structure of the matrix $\boldsymbol{M}$ is not automatically present in a generic formulation of the IBP equations. One can easily check that for the original Laporta equations, the matrix $\boldsymbol{M}$ cannot be put in such a form. This can be observed in the matrix in eq.~(\ref{eq:M}) (left).

The reader has probably noticed that we speak of the unit matrix in the top left corner, yet we emphasize it is a $1\times 1$ matrix. The reason for this is that in the general case, when there are $n$ master integrals, the above matrices will generalize to being the unit $n\times n$ unit matrix. In the examples shown above, we work in the sector $(0, 1, 0, 1)$ which has only one master, hence the $1\times 1$ matrix. 

The second important observation about eq.~(\ref{eq:M}) is that it provides us with a way to quantify the requirement for efficiency mentioned in the beginning of this section. By inspection we notice that the diagonal equation has the smallest number of non-zero elements adjacent to the main diagonal. On the other hand, the matrix-diagonal case (eq.~(\ref{eq:M}), right) has its non-zero elements located mainly two lines below the main diagonal. In light of our discussion following eq.~(\ref{eq:powers}) it is clear that a block-diagonal-lower-triangular matrix with elements that are farther away from the main diagonal can be moved to the first  $n$ rows of the matrix (in this example $n=1$) with less operations. This observation leads to the following idea: try to find a set of IBP equations which have block-diagonal-lower-triangular structure and have elements as far as possible from the main diagonal. 

In this paper we formulate one such strategy which we call IBP equations in {\it triangular form}. Before we specify the IBP equations and how one can use them, we show the top $10\times 10$ sub-matrix of $\boldsymbol{M}$ for this case:
\begin{equation}
\begin{pmatrix}
\mathbf{1} & 0 & 0 & 0 & 0 & 0 & 0 & 0 & 0 & 0 \\
{\rm x} & \mathbf{0} & 0 & 0 & 0 & 0 & 0 & 0 & 0 & 0 \\
0 & 1 & \mathbf{0} & 0 & 0 & 0 & 0 & 0 & 0 & 0 \\
0 & {\rm x} & 0 & \mathbf{0} & 0 & 0 & 0 & 0 & 0 & 0 \\
0 & {\rm x} & 0 & 1 & \mathbf{0} & 0 & 0 & 0 & 0 & 0 \\
0 & 0 & {\rm x} & 0 & 1 & \mathbf{0} & 0 & 0 & 0 & 0 \\
0 & 0 & 0 & {\rm x} & 0 & 0 & \mathbf{0} & 0 & 0 & 0 \\
0 & 0 & 0 & {\rm x} & 0 & 0 & 1 & \mathbf{0} & 0 & 0 \\
0 & 0 & 0 & 0 & 0 & 0 & 0 & 1 & \mathbf{0} & 0 \\
0 & 0 & 0 & 0 & 0 & {\rm x} & 0 & 0 & 1 & \mathbf{0}
\end{pmatrix}\,.
\label{eq:M-diag}
\end{equation}

It is evident from eq.~(\ref{eq:M-diag}) that most of the non-trivial elements are indeed located farther away from the main diagonal compared to the diagonal or matrix-diagonal approaches. As anticipated in the discussion in the beginning of this section, these equations are not diagonal anymore but are more efficient to evaluate numerically for two reasons: first, the equations are strictly lower triangular which means that in the course of their solving with any existing IBP solving program, they never require any back-substitutions. Second, the rational coefficients appearing in this approach are smaller in size than in the diagonal or matrix-diagonal approaches which makes their numeric evaluation more efficient.

\subsection{Equations in triangular form}\label{sec:triangular-eq}

In this section we present the triangular approach. As implied in the previous subsection, it produces a large system of triangular equations, which contains integrals with no abstract indices. This system of equations can then be solved by any IBP solving program.

Instead of deriving such a massive system of equations directly, it is more efficient to follow a different path: we derive a relatively small system of equations involving integrals with abstract indices, which is then used to generate the system of explicit equations by providing specific integer values for the abstract indices. In the rest of this work, when discussing the triangular approach we will have in mind this abstract set of equations. 

All abstract indices $\nu_i$ appearing in the triangular approach are assumed to take non-positive integer values, in contrast to the diagonal approach where abstract indices are not necessarily integer-valued. 

As we will explain in more detail in the next section, the indices of all integrals can be split into three groups: abstract indices, explicit positive indices and explicit non-positive ones. Abstract indices will always be considered to be non-positive integers. For this reason, for each integral we can unambiguously assign a sector based on its positive indices.

It is conceptually straightforward to formulate the triangular approach. A set of equations needs to be derived for each sector. The algorithm then maps any integral in a given sector to:
\begin{itemize}
\item Strictly {\it simpler} integrals in the same sector, or 
\item Integrals that belong to a sub-sector (which are automatically simpler). 
\end{itemize}

To fully define the approach, let us next specify the criterion for an integral in the same sector to be simpler. To that end the following analogy is very helpful: one can picture all integrals in the given sector as being located on layers inside a multidimensional {\it ball}. The dimension of each ball equals the number of non-positive indices, i.e. all abstract indices plus all explicit non-positive ones. 

To be able to strictly order any two integrals $I_a$ and $I_b$, we introduce two measures of ordering on that ball:
\begin{equation}
w(I_a) ~~ {\rm and} ~~ \Delta(I_a,I_b)\,.
\label{eq:def-distance}
\end{equation}

The parameter $w$ plays the role of a radial distance from the center of the ball, with each subsequent outer layer containing integrals of increasing $w$. Following Laporta~\cite{Laporta:2000dsw} we take $w$ to be the negative weight of an integral, which is defined as the negative sum of all its non-positive indices. In the calculation of $w$ any abstract indices are set to zero. For example, $w(I_{\nu_1-1,1,-1,1}) = 2$. This is done for convenience, ensuring that the weight is a pure number.

The second parameter, $\Delta$, introduces a strict ordering on a given layer. Continuing with our ball analogy, we define an axis which we call the {\it North pole}. $\Delta(I_a,I_b)$ is then an ordering parameter which tells us which one of its two arguments, $I_a$ or $I_b$ is closer to the North pole. 
The integral on any given layer which represents the North pole is special since it is the simplest one on this layer.

In practice, we select one of the indices $\nu_i$ to represent the North pole. 

The ordering parameter $\Delta(I_a,I_b)$ is defined as follows. We first introduce a hierarchy between the indices $\nu_i$. For example, in a problem with three indices $\nu_{1,2,3}$, we could define the North pole to be $\nu_1$ and then postulate that $\nu_2$ is closer to $\nu_1$ than $\nu_3$ is to $\nu_1$. Clearly, this ordering is a matter of choice and any other ordering between indices is allowed a priori. 

We start by comparing the farthest index. If $I_a$'s index is larger than that of $I_b$, then the integral $I_a$ is closer to the North pole. If both integrals have equal values for this index, then the next-to-farthest index is compared, and so on.

We can now state the simplicity criterion in the first bullet above as follows: an integral is simpler if
\begin{itemize}
\item It belongs to an inner layer (i.e. has smaller $w$), or 
\item It belongs to the same layer but is closer to the North pole (as determined by $\Delta$)\,.
\end{itemize}

The above imply that the integral located on the North pole on a given layer can only be mapped to integrals on an inner layer. 

One may wonder how master integrals are treated in the course of the mappings specified above. Masters are not encountered while we work with abstract indices. Master integrals appear only once the abstract triangular equations are made explicit, in the course of their solving, by providing explicit values for the indices. At this stage no mapping is done anymore. We will show an example of this procedure below.

We should point out that with this convention, the smallest value the weight $w$ can take is $w=-1$. This follows from our choice of treating abstract indices and from the following property of the abstract IBP equations: upon differentiation, see for example eq.~(\ref{eq:I}), each abstract integral can have at most one propagator raised to a power $\nu+1$. As a consequence, the integrals with $w=-1$ will always be multiplied by a factor of $\nu$. This is a very important result, because it prevents the appearance in the point $\nu=0$ of integrals belonging to a supersector (which would be at variance with the ordering procedure discussed above).

While any ordering definition of $\Delta(I_a,I_b)$ would lead to the same IBP solution, through extensive testing we have observed that different orderings lead to significant differences in the derived triangular equations. This, in turn, impacts how easily the equations can be solved. We have observed that simpler equations, i.e. ones that have smaller size, tend to be easier to solve. With this in mind, in every problem we have solved, we have chosen the hierarchy between the indices by trial and error and by selecting the one that leads to smaller equations. While such a trial and error approach leads to extra work, the derivation of the equations is relatively fast and therefore the cost for doing this is not significant. 

Before closing this section, let us demonstrate this algorithm for the case of the one-loop massless box. As explained above, unlike the case of the diagonal and matrix-diagonal equations which can be derived and solved for a group of sectors at a time, we derive and solve the triangular equations one sector at a time. 

We begin with one of the two bottom sectors, $(0, 1, 0, 1)$, and restrict $\nu_1\le0$ and $\nu_3\le0$. We choose $\nu_1$ to be the North pole. As explained above we derive an equation that maps an integral to simpler integrals in the same sector:
\begin{equation}
I_{\nu_1,1,\nu_3-1,1}=I_{\nu_1-1,1,\nu_3,1}-\frac{\nu_1-\nu_3}{2-d+\nu_1+\nu_3}I_{\nu_1,1,\nu_3,1}\,.
\label{eq:diag-1}
\end{equation}

The first term in the RHS of eq.~(\ref{eq:diag-1}) above is an integral which belongs to the same layer as the original integral but is closer to the North pole, while the second term belongs to an inner layer. Note that the RHS contains no integrals that belong to subsectors, since in this case all subsectors are zero sectors.

Finally, we need an equation which maps the integral on the North pole towards lower layers:
\begin{equation}
I_{\nu_1-1,1,0,1}=-\frac{(2 - d + 2 \nu_1)  t }{2 (2 - d + \nu_1)}I_{\nu_1,1,0,1}\,.
\label{eq:diag-2}
\end{equation}
The above two equations reduce all integrals in this sector to the master integral $I_{0,1,0,1}$.

Next we consider the sector $(1,1,0,1)$ with $\nu_3\le0$. Unlike the above two equations, this sector has a non-zero subsector, therefore we need an equation which maps integrals either towards an inner layer or to a lower sector:
\begin{equation}
I_{1,1,\nu_3-1,1} = I_{0,1,\nu_3,1}-\frac{1 - \nu_3}{3 - d + \nu_3}I_{1,1,\nu_3,1}\,.
\label{eq:diag-3}
\end{equation}

Notice that in the mapping to lower sectors we require the integrals to have smaller weight. This is the reason why the RHS of eq.~(\ref{eq:diag-3}) contains $I_{0,1,\nu_3,1}$ but not $I_{-1,1,\nu_3,1}$. The reason behind this requirement is increased efficiency: it has been noticed many times before \cite{Lange:2025fba,Smirnov:2025prc} that integrals with the highest numerator powers are only present in the top topology and not in any sub-topology. The requirement mentioned here ensures that such top integrals will not be mapped to integrals that are not already present in the amplitude of interest. 

Eq.~(\ref{eq:diag-3}) reduces any integral in this sector to the integral $I_{1,1,0,1}$, which is the integral with smallest weight $w$ in this sector (plus integrals in the lower sector which was already solved). If the integral $I_{1,1,0,1}$ is a master, then this sector is fully solved. If it is not a master, then one needs to find an additional relation for this integral to the lower sector. Such an equation can be easily derived:
\begin{equation}
I_{1,1,0,1}=-\frac{2  (3 - d) }{(4 - d) t} I_{0,1,0,1}\,.
\label{eq:diag-4}
\end{equation}

In more complicated topologies, the RHS of eq.~(\ref{eq:diag-4}) can in principle contain any integral in the lower sector(s) that has no square propagators.

The second higher sector $(0,1,1,1)$ can be treated in complete analogy. The domain is now $\nu_1\le0$ and since $I_{0,1,1,1}$ is not a master, the corresponding equations are
\begin{equation}
I_{\nu_1-1,1,1,1}=I_{\nu_1,1,0,1}-\frac{1 - \nu_1}{3 - d + \nu_1}I_{\nu_1,1,1,1}\,,
\label{eq:diag-5}
\end{equation}
and 
\begin{equation}
I_{0,1,1,1}=-\frac{2  (3 - d) }{(4 - d) t} I_{0,1,0,1}\,.
\label{eq:diag-6}
\end{equation}

To complete the calculation of the topology, one needs to repeat the above approach to the second bottom sector $(1, 0, 1, 0)$ and its two supersectors $(1, 1, 1, 0)$ and $(1, 0, 1, 1)$.

\section{The algorithm}\label{sec:algo}

As discussed in the previous sections, we have established ways for casting the IBP equations in diagonal, matrix-diagonal and triangular forms. In the previous section we demonstrated through simple examples how these three approaches work. In this section we present their derivations in the general case.

The first step in specifying the algorithm is to recognize that all three types of equations have the following common form
\begin{equation}
I_0 = c_1I_1 + c_2 I_2 + \ldots +  c_n I_n \,,
\label{eq:alg}
\end{equation}
where $I_0, I_1,\dots , I_n$ are integrals and $c_1,\dots ,c_n$ are rational coefficients. Eq.~(\ref{eq:alg}) is, essentially, an ansatz. In the following we explain how to specify the set of integrals $I_0, I_1,\dots , I_n$ for each one of the three types of equations. We then specify how to determine the rational coefficients $c_1,\dots ,c_n$. 

We would like to mention that the IBP reduction package {\tt Blade} \cite{Guan:2024byi} employs a similar idea. However, their relations are derived for fixed numerical indices, whereas ours are for abstract indices. Moreover, although our equations can also be considered as reduction rules, the algorithm used to derive them, as to be detailed in this section, does not rely on Gr$\ddot{\rm o}$bner bases \cite{Lee:2012cn}, Syzygies \cite{Smith:2025xes}, or generating functions\cite{Feng:2025leo,Feng:2026imq}.

\subsection{Determining the set of integrals}\label{sec:algo-I}

Determining the integrals $I_0, I_1,\dots , I_n$ for the case of diagonal equations is fairly straightforward. Recalling our expectation that the order $n$ of the difference equation is set by the number of non-zero master integrals $N$, we have $n=N$. In this case $I_0$ is simply $I_{\nu-n}$ and the corresponding set of integrals is $I_{\nu-n}, I_{\nu-n+1}, \dots, I_{\nu}$. Here $\nu$ is the index which is being diagonalized. The presence of other abstract indices is implicit, and they are considered fixed. 

The case of the matrix-diagonal equation is treated similarly, with the following minor adjustment. The $n=N$ integrals $I_1,\dots , I_n$ simply correspond to the $n$ elements of the vector $\boldsymbol{V}_{\nu}$. How to define the integrals that belong to the vector $\boldsymbol{V}_{\nu}$ was already discussed in sec.~\ref{sec:matrix-diag}. 

Rewriting the matrix equation as a set of $n$ scalar equations, the integral $I_0$ corresponding to each one scalar equation is the corresponding element of the shifted vector $\boldsymbol{V}_{\nu-1}$. For example, for the one-loop box case considered in sec.~\ref{sec:matrix-diag} diagonalized along the index $\nu=\nu_1$, the vector $\boldsymbol{V}_{\nu}$ contains the integrals $I_{\nu_1,1,\nu_3,1}$ and $I_{\nu_1-1,1,\nu_3,1}$ which correspond to $I_1$ and $I_2$, respectively. The integral $I_0$ for each individual scalar equation is given by the integrals $I_{\nu_1-1,1,\nu_3,1}$ and $I_{\nu_1-2,1,\nu_3,1}$, respectively. 

In the rest of this subsection we discuss the triangular equations. The main difference with respect to the two diagonal cases is that here we do not know a priori the number of integrals that enter the equation, i.e. in general we have $n\geq N$. In complicated topologies $n$ tends to be significantly larger than the number of relevant master integrals $N$. For this reason the RHS of eq.~(\ref{eq:alg}) is left to be as general as possible, subject to the following requirements.

As explained in sec.~\ref{sec:triangular-eq}, we work in a given sector. We work by deriving equations for one abstract index $\nu$ at a time, beginning with the index which is farthest from the North pole (the distance between an index and the North pole is introduced in sec.~\ref{sec:triangular-eq}). We specify the integral $I_0$ to be the generic integral belonging to this sector, which is shifted by $-1$ in the current index $\nu$. 

For example, in the one-loop case with sector $(0, 1, 0, 1)$ considered in sec.~\ref{sec:triangular-eq}, the general integral belonging to this sector reads $I_{\nu_1,1, \nu_3,1}$, and the North pole was chosen along the index $\nu_1$. Therefore, in the two resulting equations the integral $I_0$ is taken to be $I_{\nu_1,1, \nu_3-1,1}$ and $I_{\nu_1-1,1,0,1}$, respectively.

In some cases, however, the shift by $-1$ may not be sufficient. In general we need to first check if the integral corresponding to $I_0$ with all indices $\nu_i$ set to zero corresponds to a master integral. If this integral happens to be a master, then we shift the index $\nu$ with $-2$ and so on until the integral $I_0$ does not correspond to a master integral. 

Furthermore, in some cases this shift may turn out to not be sufficient to find a system of equations in a closed form (how we know if this is the case is explained in sec.~\ref{sec:implementation} below). If this happens, one needs to increase the shift in the index $\nu$ by another unit. Eventually such an equation will be found since we know that the diagonal equation exists, and this gives an upper limit for the value of this shift. Basically we are trying to find the smallest shift for which an equation can be derived. We have noticed from experience that even in the most complicated topologies, performing such checks is fairly quick.

Finally, one needs to bear in mind that in cases where an index $\nu$ needs to be shifted by $-k$ units, $k>1$, during the solving of the subsequent indices the index $\nu$ will take independently values between $0,\dots,-k+1$. For each of these values of $\nu$, a separate equation needs to be derived.

The integrals $I_1, \dots , I_n$ appearing in the RHS of eq.~(\ref{eq:alg}) could be any integral $I$ that has no squared propagators and satisfies:
\begin{enumerate}
\item $I$ is in a lower sector than $I_0$'s, with $w(I) < w(I_0)$ if $I$ is in the sector immediately below that of $I_0$, or $w(I) < w(I_0)-1$ if it lies two sectors below,  and so on. The weight $w$ is defined in eq.~(\ref{eq:def-distance}).
\item $I$ is in the same sector as $I_0$, and $w(I) < w(I_0)$.
\item $I$ is in the same sector as $I_0$,  $w(I) = w(I_0)$, and $I$ is strictly closer to the North pole than $I_0$ is, as measured by $\Delta(I,I_0)$ defined in eq.~(\ref{eq:def-distance}).
\end{enumerate}

As already discussed in sec.~\ref{sec:triangular-eq}, the restriction on the weight in point 1. above is introduced for efficiency.

\subsection{Determining the coefficients}\label{sec:algo-c}

In this subsection, we describe the algorithm used to determine the coefficients $c_1,\dots ,c_n$ from the original IBP equations. Consider the complete set of original IBP equations:
\begin{equation}
E_i(\nu_1, \nu_2, \ldots, \nu_{N_p}) = 0\,,
\label{eq:IBP}
\end{equation}
where $i$ runs from $1$ to $N_{\text{IBP}}$, the total number of IBP equations for the topology we are interested in, and $N_p$ denotes the total number of propagators. For example, in the case of the one-loop box, the full set of IBP equations is given in eqs.~(\ref{eq:IBP-1-loop-box-1},\ref{eq:IBP-1-loop-box-2},\ref{eq:IBP-1-loop-box-3},\ref{eq:IBP-1-loop-box-4}). 

The key step in the derivation of the desired equations (\ref{eq:alg}) is the realization that one has to utilize not just the IBP equations eq.~(\ref{eq:IBP}) but also their versions shifted by various integer values. Let us introduce a set of $N_p$ integers $s_i$, and recast the IBP equations $E_i(\nu_1, \nu_2, \ldots, \nu_{N_p})$ as
\begin{equation}
E_i(\underbrace{\nu_1+s_1,\dots,\nu_k+s_k}_{\text{abstract}},\underbrace{s_{k+1},\dots,s_r}_{\text{positive}},\underbrace{s_{r+1},\dots,s_{N_P}}_{\text{non-positive}})\,,
\label{eq:E-shifted}
\end{equation}
by first shifting the abstract indices $\nu_i$ by integer values, $\nu_i \to \nu_i+s_i$, and then setting some of the abstract indices to zero. The second step is needed when we need to derive equations for a given sector which is a proper subset of the full topology. In such a case we keep as abstract only the indices that contribute to the mapping eq.~(\ref{eq:alg}).

The values of $k$ and $r$ are fixed by the choice of $I_0$, i.e. it is $I_0$ that tells us which indices remain abstract and which fully numeric: all indices before $k$ are the ones that we wish to keep abstract; the indices from $k+1$ to $r$ are positive for $I_0$ and specify its sector, while the remaining indices ${r+1},\dots, N_P$ are explicit non-positive integers.

The ordering of the indices in eq.~(\ref{eq:E-shifted}) above is merely a matter of convenience. It does not affect the generality of our argument. An example of this situation is discussed in sec.~\ref{sec:reduce-order} where we derive diagonal equations for the integrals $I_{1, \nu_2, 1, \nu_4}$ belonging to the sector $(1, 0, 1, 0)$ with $N_p=4, k=2$ and $r=4$. 

We next consider a general linear combination of all possible shifted equations:
\begin{eqnarray}
&& \sum_{i=1}^{N_{\text{IBP}}}\sum_{s_1=-\infty}^{+\infty}\dots \sum_{s_{N_p}=-\infty}^{+\infty}
x_{i,s_1,\ldots,s_{N_p}} \times\nonumber\\
&& E_i(\nu_1+s_1,\dots,s_{k+1},\dots,s_{r+1},\dots,s_{N_P}) = 0\,,
\label{eq:sum-E}
\end{eqnarray}
where $x_{i,s_1,\ldots,s_{N_p}}$ are a piori unknown coefficients, which we collectively denote as $\boldsymbol{x}$. A suitable choice of these coefficients yields the desired equation (\ref{eq:alg}). 

We should point out that, as explained in detail in sec.~\ref{sec:implementation} below, the upper summation limits of $s_1,\dots,s_k$ for the triangular equations will never exceed zero. This is consistent with our discussion in sec.~\ref{sec:triangular-eq}.

Since each $E_i$ is a linear combination of integrals, eq.~(\ref{eq:sum-E}) can be rewritten as a sum over integrals:
\begin{eqnarray}
&&\sum_{s_1=-\infty}^{+\infty}\dots \sum_{s_{N_p}=-\infty}^{+\infty}
c_{s_1,\dots,s_{N_p}}(\boldsymbol{x})\times \nonumber\\
&&I_{\nu_1+s_1,\dots,s_{k+1},\dots,s_{r+1},\dots,s_{N_P}} = 0\,.
\label{eq:sum-I}
\end{eqnarray}

The required mapping eq.~(\ref{eq:alg}) is then obtained by selecting the coefficients $\boldsymbol{x}$ such that
\begin{equation}
c_{s_1,s_2,\ldots,s_{N_p}}(\boldsymbol{x}) = 
\begin{cases}
1, & \text{if } I_{\nu_1+s_1, \dots} = I_{0}, \\[6pt]
0, & \text{if } I_{\nu_1+s_1, \dots} \notin \{I_0,I_1, \ldots, I_n\}\,.
\end{cases}
\label{eq:c}
\end{equation}

Eq.~(\ref{eq:c}) above simply states that we normalize the coefficient of $I_0$ in order to exclude trivial solutions, and that we eliminate all integrals that do not belong to the ansatz eq.~(\ref{eq:alg}).

While the above construction is conceptually sound, in practice it is neither efficient nor possible to construct a linear combination out of infinite number of equations. Hence, in practice, we implement the algorithm iteratively, by working with a finite subset of all shifted equations $E_i$: we start with no equations; then we add one equation at a time and test after each new addition if it is possible to find a solution like eq.~(\ref{eq:c}). This procedure is repeated until a solution to eq.~(\ref{eq:c}) is found. The order of adding the equations is a matter of efficiency and we discuss this in sec.~\ref{sec:implementation} below.

The above strategy is implemented as follows. We denote by $\boldsymbol{e}$ the finite set of equations $e_i$ which are being tested at a given point in time. Although the equations $e_i$ are simply a subset of the infinite set of equations $E_i$, we denote them using a small letter $e$ in order to emphasize that they belong to the finite set of equations being tested. We also want to stress that the equations $e_i$ are not summed up like in eq.~(\ref{eq:sum-E}) but are kept as a list $\boldsymbol{e}$ which can be thought of as a ``vector" with elements $e_i$.

To check whether after a given step the system of equations $e_i=0$ provides enough information to find a solution in the form of eq.~(\ref{eq:c}), we rewrite the system of equations $\boldsymbol{e}=\boldsymbol{0}$ in the following way:
\begin{equation}
\sum_{j} \varepsilon_{ij} F_j = 0\,,
\label{eq:alg-B-F}
\end{equation}
where $\boldsymbol{F}$ is a vector which consists of all the integrals appearing in $\boldsymbol{e}$, and $\varepsilon_{ij}$ are coefficients defined by $e_i = \sum_j \varepsilon_{ij} F_j$. We order the integrals within the vector $\boldsymbol{F}$ as follows
\begin{equation}
\boldsymbol{F} =
\big(
\underbrace{\hat{I}_1, \hat{I}_2, \hat{I}_3, \ldots, \hat{I}_{n_{\rm out}}}_{\text{not in eq.}\,(\ref{eq:alg})},
I_0, I_1, \ldots, I_n \big)^T \,.
\label{eq:alg-F}
\end{equation}

Hence, the components $\varepsilon_{ij}$ with $1 \le j \le n_{\rm out}$ correspond to the coefficients of the integrals that do not belong to the mapping. We denote these by $\varepsilon^{(1)}_{ij}$. Notice that $\varepsilon^{(1)}_{ij}$ is rectangular, and $\text{dim}(\varepsilon^{(1)}_{ij})$ is defined as its smaller dimension which is normally the number of rows. 

If there exists a linear combination of $e_i$ such that the coefficients of all integrals that do not belong to the desired equation eq.~(\ref{eq:alg}) vanish, we find that
\begin{equation}
\text{Rank}\left(\varepsilon^{(1)}_{ij}\right) < \text{dim}\left(\varepsilon^{(1)}_{ij}\right)\,.
\label{eq:alg-eps1}
\end{equation}
This condition implies that the set of equations $e_i$ are linearly dependent with respect to the integrals that do not belong to the mapping. Consequently, there exists a linear combination that eliminates all these integrals, which is equivalent to finding a non-trivial solution for the second line of eq.~(\ref{eq:c}).

Next, we must ensure that the obtained solution also satisfies the first line of eq.~(\ref{eq:c}). To verify this, we consider a slightly augmented matrix $\varepsilon^{(2)}_{ij}$ defined by $\varepsilon_{ij}$ with $1 \le j \le n_{\rm out} + 1$, i.e. we also include the coefficient of $I_0$.

Given that eq.~(\ref{eq:alg-eps1}) holds, if we further require that the linear combination does not remove $I_0$, the following condition must be satisfied:
\begin{equation}
\text{Rank}\left(\varepsilon^{(2)}_{ij}\right) = \text{dim}\left(\varepsilon^{(2)}_{ij}\right)\,.
\label{eq:alg-eps2}
\end{equation}

This condition ensures that when more integrals are included, all $e_i$ remain linearly independent in this extended space. If eq.~(\ref{eq:alg-eps2}) is not satisfied, the linear combination found earlier that led to eq.~(\ref{eq:alg-eps1}) is not a valid one and cannot be used to determine eq.~(\ref{eq:alg}). In this case, we must discard the last added equation from the list $\boldsymbol{e}$ and proceed to the next iteration. 

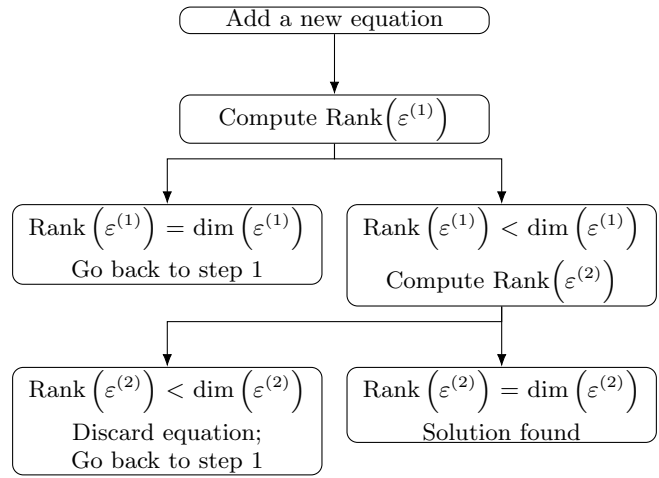
\begin{figure}[t]
\centering
\begin{tikzpicture}[
    box/.style={
        rectangle, draw, rounded corners,
        text width=4cm,
        align=center,
        inner sep=1pt
    },
    node distance=4mm,
    >={Latex}
]
\node[box] (new) {Add a new equation};
\node[box, below=8mm of new] (r1)
{
    Compute Rank$\left(\varepsilon^{(1)}\right)$
};
\node[box, below=8mm of r1, xshift=-22mm] (cont)
{
    ${\rm Rank}\left(\varepsilon^{(1)}\right)$ = $\dim\left(\varepsilon^{(1)}\right)$ \\[1mm]
    Go back to step 1
};
\node[box, below=8mm of r1, xshift=22mm] (r2)
{
    ${\rm Rank}\left(\varepsilon^{(1)}\right)$ $<$ $\dim\left(\varepsilon^{(1)}\right)$ \\[1mm]
    Compute Rank$\left(\varepsilon^{(2)}\right)$
};
\node[box, below=8mm of r2, xshift=-44mm] (discard)
{
    ${\rm Rank}\left(\varepsilon^{(2)}\right)$ $<$ $\dim\left(\varepsilon^{(2)}\right)$ \\[1mm]
   Discard equation; \\ Go back to step 1 
};
\node[box, below=8mm of r2, xshift=0mm] (sol)
{
    ${\rm Rank}\left(\varepsilon^{(2)}\right)$ = $\dim\left(\varepsilon^{(2)}\right)$ \\[1mm]
    Solution found
};
\draw[->] (new) -- (r1);
\draw[->] (r1.south) --++(0,-2mm) --++(-22mm,0) -- (cont.north);
\draw[->] (r1.south) --++(0,-2mm) --++(22mm,0) -- (r2.north);
\draw[->] (r2.south) --++(0,-2mm) --++(-44mm,0) -- (discard.north);
\draw[->] (r2.south) --++(0,-2mm) --++(0mm,0) -- (sol.north);
\end{tikzpicture}
\caption{Flowchart of the rank-based decision process.}
\label{fig:chart}
\end{figure}

The procedure is summarized in the flow chart in fig.~\ref{fig:chart}. Among all topologies computed in this work, see sec.~\ref{sec:benchmarks}, the maximum rank we have encountered is 3210. We have established that even with matrices of such size, the rank can be computed efficiently if we replace all variables with integer values.

Up to this point we showed how to find a set of shifted IBP equations that satisfy eq.~(\ref{eq:c}). Our procedure used the calculation of the rank of matrices as an indicator for the existence of this solution. However we still need to find the actual linear combination of IBP equations that leads to eq.~(\ref{eq:c}). 

This is a straightforward task, that can be accomplished following the steps outlined around eqs.~(\ref{eq:sum-E},\ref{eq:sum-I},\ref{eq:c}). The only difference is that instead of the infinite system of equations $E_i$, we use the finite set of equations $e_i$ constructed in the steps above. We already showed that for the set of equations $e_i$ a unique solution exists, which can be found using direct linear equations methods. In practice, we do not need the coefficients $\boldsymbol{x}$ but directly the coefficients $c_{s_1,\dots,s_{N_p}}$ appearing in eq.~(\ref{eq:sum-I}). They can be determined by matching eq.~(\ref{eq:sum-I}) to eq.~(\ref{eq:c}), which can be performed very efficiently with the help of finite field methods.

\subsection{Implementation details}\label{sec:implementation}

In the previous subsection we explained that the desired equations can be obtained from a finite subset of all possible shifted IBP equations. In the following we explain how to (efficiently) select this subset of equations.

We start the generation of equations from the following point in the space of shifts $s_i$
\begin{equation}
\boldsymbol{\hat{s}}=\{\underbrace{0,\dots,0}_{1\;\text{to}\;k},\underbrace{1,\dots,1}_{k+1\;\text{to}\;r},\underbrace{0,\dots,0}_{r+1\;\text{to}\;N_P}\} \,.
\label{eq:begin-s}
\end{equation}

We then generate equations $e_i$ by applying shifts $s_i$, starting from the point $\boldsymbol{\hat{s}}$ and requiring $s_i\leq \hat{s}_i$ for all indices. This restriction ensures that in the case of the triangular equations the abstract indices are never shifted towards positive values for any $\nu_i\leq 0$. As discussed above, the only exception are integrals of the type $I_{\dots,\nu+1,\dots}$, however, {\it in IBP equations} they are always multiplied by a factor of $\nu$ which prevents the appearance of integrals with negative weight.

The  constraint $s_i\leq \hat{s}_i$ is not strictly required in the case of diagonal equations. It is nonetheless beneficial in this case as well, since positive shifts do not contribute to the derivation of diagonal equations. 

The shifts on the ``positive" and ``non-positive" indices are constrained for the same reason: improved efficiency of equation generation.

The shifts are ordered in terms of the parameter
\begin{equation}
\delta=\sum_{i=1}^{N_P}(\hat{s}_i-s_i)\,.
\label{eq:def-weight}
\end{equation}

We first generate all equations with $\delta=0$, then the ones with $\delta=1$, etc., until we reach the value $\delta=w(I_0)+1$. We have commonly observed that at this point the system closes. From experience we have also established that if a system does not close at this point, it likely will not close at all. This can happen when the attempted ansatz for the system of equations is too restrictive, see sec.~\ref{sec:algo-I} for a description of when this may happen and how to deal with such situations.

Finally, we note that we do not generate the equations from zero sectors since they only include zero integrals. The triangular approach can be further optimized and produce equations with smaller size, if one applies the above set of equations in a slightly different order, by prioritizing the increase in the index which is being shifted in $I_0$.

\section{Benchmark results for the triangular approach}\label{sec:benchmarks}

\begin{figure}[t]
\centering
\begin{tikzpicture}[scale=0.3]
\node (p1) at  (-3.5,1.3)  {$p_1$};
\node (p2) at  (-3.5,-2.7) {$p_2$};
\node (p3) at  (3.5,1.3)   {$p_3$};
\node (p4) at  (3.5,-2.7)  {$p_4$};
\draw  (-4,2) to (4,2) {};
\draw  (-4,-2) to (4,-2) {};
\draw  (-2,-2) to (0,2) {};
\draw  (-2,2) to (0,-2) {};
\draw  (2,-2) to (2,2) {};
\node (dummyleft) at (-4.2,0) { };
\node (dummyright) at (4.2,0) { };
\end{tikzpicture}
\begin{tikzpicture}[scale=0.3]
\node (p1) at  (-3.5,1.3)  {$p_1$};
\node (p2) at  (-3.5,-2.7) {$p_2$};
\node (p3) at  (3.5,1.3)   {$p_3$};
\node (p4) at  (3.5,-2.7)  {$p_4$};
\node (p5) at  (3.5,-0.7)  {$p_5$};
\draw  (-4,2) to (4,2) {};
\draw  (-4,-2) to (4,-2) {};
\draw  (-2,-2) to (-2,2) {};
\draw  (2,-2) to (2,2) {};
\draw  (0,-2) to (0,2) {};
\draw  (2,0) to (4,0) {};
\node (dummyleft) at (-4.2,0) { };
\node (dummyright) at (4.2,0) { };
\end{tikzpicture}
\begin{tikzpicture}[scale=0.3]
\node (p1) at  (-3.5,1.3)  {$p_1$};
\node (p2) at  (-3.5,-2.7) {$p_2$};
\node (p3) at  (3.5,1.3)   {$p_3$};
\node (p4) at  (3.5,-2.7)  {$p_4$};
\node (p5) at  (1.0,-0.7)  {$p_5$};
\draw  (-4,2) to (4,2) {};
\draw  (-4,-2) to (4,-2) {};
\draw  (-2,-2) to (-2,2) {};
\draw  (2,-2) to (2,2) {};
\draw  (0,-2) to (0,2) {};
\draw  (0,0) to (1.5,0) {};
\node (dummyleft) at (-4.2,0) { };
\node (dummyright) at (4.2,0) { };
\end{tikzpicture}
\caption{Two-loop topologies: off-shell di-boson (left), three jet planar $C_1$ (center) and three jet non-planar $B_1$ (right).}
\label{fig:topologies}
\end{figure}
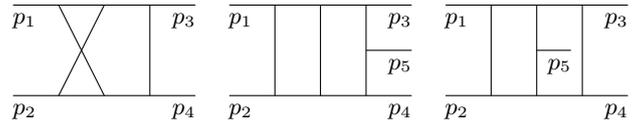

In this section we present results for several non-trivial calculations performed with our triangular algorithm. We perform a tuned comparison between this algorithm and the syzygy-based algorithm as a representative of the current state-of-the-art. This has allowed us to verify the correctness of our algorithm in all cases we have considered and to study in detail its performance. 

Specifically, we test our approach using three two-loop topologies. The simplest one is the non-planar amplitude for off-shell di-boson production, which is shown in fig.~\ref{fig:topologies} (left) with $p_1^2=p_2^2=0$ and $p_3^2\neq p_4^2 \neq 0$. We next consider two topologies appearing in three jet production: the planar topology $C_1$  shown in fig.~\ref{fig:topologies} (center) and the most complicated non-planar topology $B_1$  shown in fig.~\ref{fig:topologies} (right). The explicit definitions for these topologies can be found in ref.~\cite{Chawdhry:2018awn}.

We use {\tt Kira 3.0}~\cite{Lange:2025fba} to carry out the reductions. In all calculations that use finite field reconstruction, we have utilized the finite field library {\tt FireFly}~\cite{Klappert:2019emp,Klappert:2020aqs} with internal optimization {\tt bunch\_size=128}.

For each topology, the system is solved up to $-5$ in the top sector, up to $-4$ in the next lower sectors, up to $-3$ in the subsequent ones, and so on. The set of integrals we reduce includes the most complicated integrals contributing to 5-gluon amplitudes.

All syzygy equations are derived using {\tt NeatIBP}~\cite{Wu:2023upw} with the spanning-cut technique. The timings reported for the syzygy solutions are summed over all spanning cuts. For each spanning cut, we simultaneously solve all master integrals within the cut. Some master integrals appear in multiple spanning cuts; in such cases, each master is solved only once -- the first time it appears -- and is set to zero when encountered again in a different cut. For the triangular equations, we derive them without setting any non-trivial sector to zero. Thus, the equations solve all integrals and all masters simultaneously when provided to the solver. For the $B_1$ topology, we only solved two spanning cuts, as the computation is particularly time-consuming. Specifically we computed the spanning cut above the master $M_6$ (defined as $I_{0,0,1,0,1,0,1,0,0,0,0}$), denoted as $\overline{M_6}$, which contains 47 masters including $M_6$, as well as the spanning cut above the master $M_{10}$ (defined as $I_{1,0,0,0,0,0,1,1,0,0,0}$) which contains 41 masters including $M_{10}$.

We have fully solved the $B_1$ topology for a rational kinematic point where only the space-time dimension has been kept abstract. 

For our triangular algorithm we have also utilized {\tt Kira}'s ability to use the library {\tt Fermat}~\cite{Fermat} as a way of directly solving the IBP equations, without relying on finite field reconstruction. 

The set of triangular IBP equations used in this section are available for download from the website \cite{website}.

All computations are performed on a standard computing server Viglen Intel S2600WF (2 x 12-core Xeon Silver 4116) with 512GB memory. The memory consumption for all jobs has been fairly low, below 150 GB, which is well within the server's available memory. Interestingly, we have noticed that the memory consumption with {\tt Fermat} is about a factor of two smaller than when using finite field reconstruction.

\begin{table}[t]
\centering
\renewcommand{\arraystretch}{1.2}
\begin{tabular}{l|c|l|c|c |c c}
\hline
\textbf{topology} & \textbf{parallelise} &
\textbf{Fermat} & 
\multicolumn{3}{c}{\textbf{Finite field}}  \\ 
\cline{3-6}
 & (cores) & triang. & syzygy & triang. & Laporta\\ 
\hline
di-boson & 24 & 470 s & 537 s & 678 s & 2216 s \\[2pt]
\hline
$C_1$ & 24 & 3.4 h & 11.3 h & 8.0 h & \\[2pt]
 & 1 & 9.4 h & 161.9 h & 135.9 h & \\[2pt]
 \hline
$B_1$: $\overline{M_6}$ & 24 & 2.31 d & 3.21 d & 5.41 d & \\[2pt]
$B_1$: $\overline{M_{10}}$ & 24 & 1.38 d & 3.51 d & 2.82 d & \\[2pt]
\hline
$B_1$ (rational & 24 & 41 s & 590.5 s & 622.2 s & \\[2pt]
kin. point) & 1 & 187.4 s & 1657.4 s & 994.7 s & \\
\hline
\end{tabular}
\caption{Performance comparison for different topologies and configurations.}
\label{tab:results}
\end{table}

In table~\ref{tab:results} we show the timings for all computations. The main feature to note is that the timings using finite field reconstruction with our triangular method and with the syzygies are comparable for all calculations, with differences never exceeding a factor of two. Neither approach is consistently faster, and each one shows better performance in some scenarios. We also note that the two approaches do not benefit from parallelization the same way, although no clear pattern emerges. 

A notable feature is the performance of our triangular algorithm using {\tt Kira}'s implementation of the {\tt Fermat} library \cite{Maierhofer:2017gsa}. In all calculations we have performed this was the fastest performing approach, outperforming the finite field calculations with between 50\% and a factor of 15. The calculations with {\tt Fermat} do not benefit as much from the available parallelization as the calculations based on finite fields do.

For completeness, we have also tested the performance of syzygy equations using {\tt Fermat} for the off-shell di-boson topology. We adopted the same setup as that used for the processes reported in table \ref{tab:results}. We found that the reduction takes 4.8 hours to complete, which is approximately 27 times slower than the its performance with finite field reconstruction and also significantly slower than the performance of our equations with either technique.

We have also incorporated an alternative linear solver with finite field reconstruction {\tt Ratracer} \cite{Magerya:2022hvj} into our framework and as an example computed the massless planar 5-point topology $C_1$. We adopted the same setup as that used for the processes reported in table \ref{tab:results}, with the level of parallelization set to 24. With the help of {\tt Ratracer}, the time required to reduce the triangular equations improved from 8.0 hours to 3.8 hours, while the reduction time for the syzygy equations was improved from 11.3 hours to 3.9 hours. It can be seen that both methods benefit from the use of  {\tt Ratracer}.

Given our expectation that our triangular algorithm is very efficient, we would expect that it should outperform the syzygies-based one. We only observe this for the {\tt Fermat}-based calculation but not for the finite field based one. On the other hand, the fact {\tt Fermat} outperforms the finite field calculations comes as a surprise. In order to clarify this, in table~\ref{tab:perprobe} we show a breakdown of the per-probe timings for both finite field calculations in the following two topologies: di-boson and the planar five point one $C_1$. 

\begin{table}[t]
\centering
\renewcommand{\arraystretch}{1.2}
\begin{tabular}{c|c c|c c}
\hline
Step per probe & \multicolumn{2}{c|}{off-shell di-boson} & \multicolumn{2}{c}{ $C_1$} \\ \cline{2-5}& syzygy & triang. &syzygy & triang. \\
\hline
Coefficient evaluation & 8.7\% & 85.8\% & 2.1\% & 92.7\%\\
Forward elimination    & 70.9\% & 11.8\% & 88.6\% & 6.0\%\\
Back substitution      & 20.4\% & 2.4\% & 9.3\% & 1.3\%\\
total                  & 100\% & 100\% & 100\% & 100\%\\
\hline
\end{tabular}
\caption{Timing breakdown per finite-field calculation. All timings reported as a percentage out of the total per-probe time.}
\label{tab:perprobe}
\end{table}
\begin{table}[t]
\centering
\renewcommand{\arraystretch}{1.2}
\begin{tabular}{c|c c|c c}
\hline
Step & \multicolumn{2}{c|}{off-shell di-boson} & \multicolumn{2}{c}{$C_1$} \\ \cline{2-5}& syzygy & triang. &syzygy & triang. \\
\hline
black-box probe        & 1.13 h & 2.17 h & 116.18 h & 78.39 h\\
Rational reconstruction& 0.94 h & 0.99 h & 45.85 h & 51.76 h\\
Overheads              & 0.50 h & 0.85 h & 11.61 h & 25.53 h\\
total                  & 2.57 h & 4.01 h & 173.64 h & 155.68 h\\
\hline
\end{tabular}
\caption{CPU hour breakdown for finite field reconstruction.}
\label{tab:CPUh}
\end{table}

We observe that for the syzygy-based equations, almost all the time is spent in the forward-elimination step (i.e. IBP solving), whereas for our equations the dominant time spent is for the rational coefficient evaluation. This trend only becomes more pronounced as the topology becomes more complicated. For example in the calculation of $C_1$ our triangular approach spends about 93\% of the time on evaluating the rational coefficients and only about 6\% in actually solving the IBP equations. For the syzygies the picture is exactly the opposite: 89\% of the time goes for solving the IBP equations while only 2\% of the time is used for the evaluation of the rational coefficients.

From this we can conclude that, indeed, our triangular approach is very efficient in solving the IBP equations. In this regard it significantly outperforms the syzygy-based equations. It also appears that this trend becomes more pronounced for more complex topologies. 

The fact that our triangular equations are slower to evaluate numerically can be expected because their size is much larger than the syzygy equations' one. The time difference we observe scales correctly with the size of the equations. This implies that our triangular equations will tremendously benefit from speeding up their evaluation, which is the main bottleneck for this approach.

Finally, to fully understand the triangular algorithm's performance with finite field reconstruction, in table~\ref{tab:CPUh} we show the CPU times for the off-shell di-boson and $C_1$ topologies with finite field reconstruction. From this table one can see that only about 50\% of the total time is taken by the times shown in table~\ref{tab:perprobe}. The remaining time is taken by rational reconstruction (which, as expected, is roughly independent of the calculation approach) and internal {\tt Kira} overheads.

\section{Conclusions}\label{sec:conclusions}

In this work, we have demonstrated that one can diagonalize the IBP equations. The diagonalization algorithm is general and applies to any IBP system. To the best of our knowledge, this is the first time such a general solution has been proposed. We have tested it and verified its correctness in a number of nontrivial one- and two-loop topologies.

We have recast our diagonalization algorithm as a first-order matrix equation. This formulation, which we call {\it matrix diagonal}, is very efficient and seems to be well suited as a starting point for deriving solutions in analytic form.

As a by-product of our work, we have developed a criterion for the efficiency of an IBP-solving algorithm. Based on this, we propose another algorithm, which we call {\it triangular}. The equations produced with this algorithm have strictly triangular form. The triangular equations can be directly implemented in existing IBP-solving programs and have the advantage that they do not require back-substitutions. For this reason, their solving is expected to be very efficient. A second feature of this algorithm that further boosts its efficiency is that it does not lead to squared propagators.

We have tested our algorithms for several complicated two-loop topologies. The triangular algorithm is at least comparable to existing state-of-the-art approaches. It also shows strong potential for further speed-up improvements, by an order of magnitude or more, if its current bottleneck is improved: faster numerical evaluation of the triangular equations due to their larger size. For these reasons, we believe the algorithms presented in this work will be useful for solving demanding problems.

The diagonal equations proposed in this work -- especially in the matrix formulation -- are indispensable for reducing integrals with high numerator powers or when numerator powers are treated as abstract variables. Possible applications include multivariate Mellin representations for gauge theory amplitudes and cross sections. The diagonal IBP equations make it possible to derive solutions to the IBP identities in closed analytic form, which may offer deeper insight into the analytic structures underpinning the set of loop integrals. We have also successfully applied this technology to seemingly unrelated objects, such as the contiguous relations of the Gauss hypergeometric function, and shown that they behave similarly to IBP identities.
 
Before closing, we would like to point out one further potential application of the techniques developed in this work: the evaluation of loop integrals in closed form. In a Mellin-space representation the dependence of a loop integral on all kinematic variables is encoded in difference equations that descend directly from the IBP identities, see e.g. ref.~\cite{Mitov:2005ps}. With the help of the techniques developed in this work, such multivariate equations can be brought to solvable form. The only remaining variable is then the space-time dimension $d$, whose dependence is governed by yet another difference equation --- the dimensional-shift identity of Tarasov~\cite{Tarasov:1996br} --- and can be solved by exploiting the integral's analytic properties in $d$, as demonstrated in ref.~\cite{Lee:2009dh}. In such a scenario the complete functional dependence of a loop integral would be determined by difference equations alone, with boundary conditions that are pure numbers. A careful investigation of this possibility is beyond the scope of the present work and we intend to return to it in the future.

\begin{acknowledgments}
We thank Herschel Chawdhry for tests and comparisons using his private IBP software, and Fabian Lange for clarifications regarding the library {\tt FireFly}. The work of J.W.L. has been supported by the University of Cambridge Harding Distinguished Postgraduate Scholars Programme and the STFC DTP research studentship grant ST/Y509140/1. The work of A.M. has been supported by STFC consolidated HEP theory grants ST/T000694/1 and ST/X000664/1.
\end{acknowledgments}

\appendix
\section{Contiguous Relations}\label{app:F21}

The contiguous relations eqs.~(\ref{eq:F21-a},\ref{eq:F21-b},\ref{eq:F21-c}) can be treated with the methods used to solve IBP equations. We have verified that eqs.~(\ref{eq:F21-a},\ref{eq:F21-b},\ref{eq:F21-c}) form a closed system which has a finite basis solution with two ``masters", chosen as:
\begin{equation}
{}_2F_1(1,1,1,z)=\frac{1}{1-z}~,~
{}_2F_1(1,1,2,z)=-\frac{\log(1-z)}{z}\,.
\label{equ.2.2}
\end{equation}

We have checked that out of the 15 known contiguous relations \cite{BatemanHTF} only three are independent (in the sense usually used in the IBP-related literature). We have chosen to work with the relations in eqs.~(\ref{eq:F21-a},\ref{eq:F21-b},\ref{eq:F21-c}). 

Following the approach of sec.~\ref{sec:matrix-diag} we cast the contiguous relations for the Gauss hypergeometric function ${}_2F_1$ in a diagonal form. Defining:
\begin{equation}
\boldsymbol{V}(a,b,c)
=\big({}_2F_1(a,b,c,z),\,{}_2F_1(a,b,c+1,z)\big)^T\,,
\label{equ.2.1}
\end{equation}
we arrive at the following matrix equations:
\begin{eqnarray}
\boldsymbol{V}(1,1,c+1)&=&
\begin{pmatrix}
0 & 1 \\
\frac{(1+c)(1-z)}{c\, z} & \frac{(1+c)( 2c z-c - z )}{c^{2} z}
\end{pmatrix}
\boldsymbol{V}(1,1,c)\,,\nonumber\\
\boldsymbol{V}(1,b+1,c)&=&
\begin{pmatrix}
\frac{\,b + z - c z\,}{b( 1- z)} & \frac{(b - c)(1 - c)\, z}{b c( 1-z )} \\
\frac{c}{b} & \frac{b - c}{b}
\end{pmatrix}
\boldsymbol{V}(1,b,c)\,,\nonumber\\
\boldsymbol{V}(a+1,b,c)&=&
\begin{pmatrix}
\frac{a + b z - c z}{a( 1 - z)} & \frac{(a - c)(b - c)\, z}{a c( 1 - z)} \\
\frac{c}{a} & \frac{a - c}{a}
\end{pmatrix}
\boldsymbol{V}(a,b,c) \,,
\label{eq:app-F21-eqs}
\end{eqnarray}
with the domain $a,b,c\ge1$. 

One can extract relevant information about the hypergeometric functions from the diagonal equation. For example, when iterating the last equation across the point $a=c$, the matrix collapses to:
\begin{equation}
\begin{pmatrix}
\frac{c + b z - c z}{a( 1 - z)} &
0 \\
1 &
0
\end{pmatrix}\,.
\end{equation}

The vanishing of the second column means that the second boundary condition is cut, i.e. it does not contribute. Hence, one can conclude that the logarithmic contribution only applies when $a < c$. By considering the other equations, one can find that the logarithmic contribution only applies for integer $a < c$ and $b < c$.

This result is, in fact, a non-trivial property of the $_2F_1$ hypergeometric function. It can most easily be observed using the following transformation \cite{BatemanHTF}:
\begin{equation}
{}_2F_1(a,b,c,z)= (1-z)^{c-a-b}{}_2F_1(c-a,c-b,c,z)\,.
\label{eq:F21transform}
\end{equation}

For integer-valued $a\ge c$ the function $_2F_1$ on the RHS collapses to a polynomial which demonstrates that, indeed, the LHS is a rational function without any logarithmic contribution.

This example demonstrates the power of the matrix-diagonalized approach for not only solving recurrence relations but for also offering direct information into the analytic properties of the functions being solved.

\begin{figure}[t]
\centering
\begin{tikzpicture}
\begin{scope}[shift={(-2.2,0)}]
    \draw[->] (-1.75,0) -- (1.75,0) node[right] {$a$};
    \draw[->] (0,-1.75) -- (0,1.75) node[above] {$c$};
    \node[below left] at (0,0) {$(1,1)$};
    \fill[red,opacity=0.25]
        (0,0) --
        (0,1.75) --
        (1.75,1.75) --
        cycle;
    \draw[thick] (0,0) -- (1.75,1.75);
\end{scope}
\begin{scope}[shift={(2.2,0)}]
    \draw[->] (-1.75,0) -- (1.75,0) node[right] {$\nu_1$};
    \draw[->] (0,-1.75) -- (0,1.75) node[above] {$\nu_3$};
    \node[below left] at (0,0) {$(1,1)$};
    \fill[red,opacity=0.25]
        (0,0) --
        (1.75,0) --
        (1.75,1.75) --
        (0,1.75) --
        cycle;
\end{scope}
\end{tikzpicture}
\caption{The sector structure for $_2F_1$ (left) and the one-loop-box topology (right); the shaded regions are the ones with two boundary conditions and the remaining are the ones with one boundary condition (applies to integer-valued indices).}
\label{fig:sectors}
\end{figure}
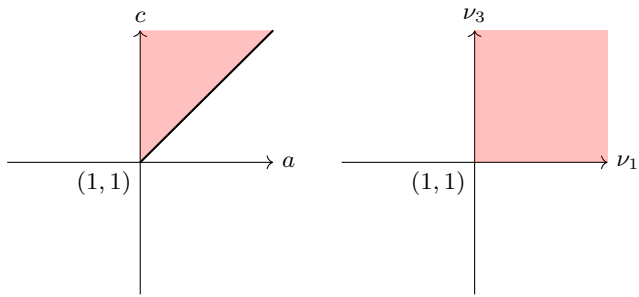

In conclusion we would like to point out the similarity between the contiguous relation and the matrix diagonal equations of the one-loop box IBP system in eqs.~(\ref{eq:W2},\ref{eq:W1}), where the zeros at $\nu_1=1$ and $\nu_3=1$ reflect the sector structure of the IBP system. The sector structure for both systems is shown in fig. \ref{fig:sectors}.

\section{Fibonacci recurrence in matrix form}\label{app:fibonacci}

Following our discussion in sec.~\ref{sec:matrix-diag}, we rewrite the Fibonacci sequence eq.~(\ref{eq:Fibonacci}) in matrix form $\boldsymbol{V}_{n+1}=\boldsymbol{W}\boldsymbol{V}_{n}$:
\begin{equation}
\begin{pmatrix} F(n+1)\\F(n+2) \end{pmatrix}
=\begin{pmatrix} 0& 1\\1&1 \end{pmatrix}\begin{pmatrix} F(n)\\F(n+1) \end{pmatrix}
\label{eq:app-fib-matrix}
\end{equation}

The matrix $ \boldsymbol{W}$ in the recurrence above can be diagonalized with the help of a similarity transformation
\begin{equation}\label{equ.1.2}
\begin{pmatrix} 0& 1\\1&1 \end{pmatrix}= \boldsymbol{S}\begin{pmatrix} \varphi & 0\\0&1-\varphi \end{pmatrix}  \boldsymbol{S}^{-1},
\end{equation}
where $\varphi=(1+\sqrt{5})/2$ was introduced in sec.~\ref{sec:intro}, and
\begin{equation}\label{equ.1.3}
 \boldsymbol{S}=\begin{pmatrix} \frac{1}{\varphi}& \frac{1}{1-\varphi}\\1&1 \end{pmatrix}.
\end{equation}

The transformation above defines a new basis $\boldsymbol{U}(n)=(G_a(n),G_b(n))^T$ such that:
\begin{equation}\label{equ.1.4}
\begin{pmatrix}F(n)\\F(n+1) \end{pmatrix}
= \boldsymbol{S}\begin{pmatrix} G_a(n)\\G_b(n) \end{pmatrix}\,.
\end{equation}

In this new basis, the recurrence relation takes the form
\begin{equation}\label{equ.1.5}
\begin{pmatrix} G_a(n+1)\\G_b(n+1) \end{pmatrix}
=\begin{pmatrix} \varphi & 0\\0&1-\varphi \end{pmatrix}\begin{pmatrix} G_a(n)\\G_b(n) \end{pmatrix}.
\end{equation}

The matrix recurrence is now decoupled and gives two first-order recurrence relations for the functions $G_a$ and $G_b$, respectively. It can be easily solved: 
\begin{equation} \label{equ.1.7}
G_a(n)=\varphi^n\;G_a(0),\quad G_b(n)=(1-\varphi)^n\;G_b(0).
\end{equation}

In light of the discussion in sec.~\ref{sec:matrix-diag} it is instructive to point out that the above solutions are particular cases of hypergeometric functions: $(1-\varphi)^n = {}_1F_0(-n,\varphi)$.

Lastly, we can reverse the transformation to obtain the solution for $F(n)$:
\begin{equation} \label{equ.1.8}
\begin{pmatrix} F(n)\\F(n+1) \end{pmatrix}
= \boldsymbol{S}\begin{pmatrix}\varphi^n&0\\0& (1-\varphi)^n\end{pmatrix} \boldsymbol{S}^{-1}\begin{pmatrix}F(0)\\F(1) \end{pmatrix}.
\end{equation}

After some simplification and rearrangement we recover the result eq.~(\ref{eq:Fibonacci-closed-form}) for $F(n)$ and its shifted $n\to n+1$ version for $F(n+1)$.

\section{The diagonal equations for the one-loop box}\label{app:diageq}

The coefficients $r_k$ appearing in eq.~(\ref{eq:diag-abstract}) read
\begin{equation}
r_0=\frac{r_{01}}{r_{02}},\quad r_1=\frac{r_{11}}{r_{12}},\quad r_2=\frac{r_{21}}{r_{22}}\,,
\label{eq:r0}
\end{equation}
where:
\begin{eqnarray}
r_{01} &=& -t(\nu_1-2) (\nu_1-1) (d - 2 \nu_1 - 2 \nu_2) (d - 2 \nu_1 - 2 \nu_4) \,,\nonumber\\
r_{02} &=& 2 (6 + d - 2\nu_{1} - 2\nu_{2} - 2\nu_{3} - 2\nu_{4}) \times\nonumber\\
&& (d - \nu_{1} - \nu_{2} - \nu_{3} - \nu_{4})\times\nonumber\\
&& (1 + d - \nu_{1} - \nu_{2} - \nu_{3} - \nu_{4}) \times\nonumber\\
&& (2 + d - \nu_{1} - \nu_{2} - \nu_{3} - \nu_{4}) \,,\nonumber
\end{eqnarray}
\vspace{0.0001em}
\begin{eqnarray}
r_{11} &=& ( \nu_{1}-2) [( \nu_{3} - \nu_{1}+1)( 2 + d-2\nu_{1} - 2\nu_{2} - 2\nu_{4})-\nonumber\\
&& t(4 + 4d + d^{2} - 8\nu_{1} - 4d\nu_{1} + 4\nu_{1}^{2} - 4\nu_{2} - 2d\nu_{2} + \nonumber\\
&& 4\nu_{1}\nu_{2} - 2\nu_{3} - 2d\nu_{3} + 4\nu_{1}\nu_{3} + 2\nu_{2}\nu_{3} - 4\nu_{4} -  \nonumber\\
&& 2d\nu_{4} + 4\nu_{1}\nu_{4} + 4\nu_{2}\nu_{4} + 2\nu_{3}\nu_{4})] \,,\nonumber\\
r_{12} &=& (6 + d - 2\nu_{1} - 2\nu_{2} - 2\nu_{3} - 2\nu_{4}) \times \nonumber\\
&& (1 + d - \nu_{1} - \nu_{2} - \nu_{3} - \nu_{4}) \times \nonumber\\
&& (2 + d - \nu_{1} - \nu_{2} - \nu_{3} - \nu_{4}) \,,\nonumber\\
r_{21}&=& 2(16 + 4d - 16\nu_{1} - 2d\nu_{1} + 4\nu_{1}^{2} - 8\nu_{2} + 4\nu_{1}\nu_{2} +  \nonumber\\
&& 2\nu_{3} + d\nu_{3} - 2\nu_{2}\nu_{3} - 2\nu_{3}^{2} - 8\nu_{4} + 4\nu_{1}\nu_{4} - 2\nu_{3}\nu_{4}) -  \nonumber\\
&& t(4 + d - 2\nu_{1} - 2\nu_{2} - 2\nu_{3})(4 + d - 2\nu_{1} - 2\nu_{3} - 2\nu_{4}) \,,\nonumber\\
r_{22} &=& 2 (6 + d - 2\nu_{1} - 2\nu_{2} - 2\nu_{3} - 2\nu_{4}) \times\nonumber\\
&& (2 + d - \nu_{1} - \nu_{2} - \nu_{3} - \nu_{4})\,.\nonumber
\end{eqnarray}

Following the discussion in sec.~\ref{sec:reduce-order}, we next present the second order diagonal equation which is restricted to the bottom group of sectors $\overline{(0, 1, 0, 1)}$. The equation in the index $\nu_1$ reads
\begin{equation}
I_{\nu_1-2,1,\nu_3,1} = \hat{r}_1 I_{\nu_1-1,1,\nu_3,1}+\hat{r}_2I_{\nu_1,1,\nu_3,1} \,,
\label{eq:olb-diag-1}
\end{equation}
where
\begin{eqnarray}
\hat{r}_1 &=& \frac{(2 - 2 \nu_1 + 2 \nu_3 - d t + 2 \nu_1 t + 2 \nu_3 t)}{2 (-1 + d - \nu_1 - \nu_3)} \,,\nonumber\\
\hat{r}_2 &=& \frac{(2 - d + 2 \nu_1)  ( \nu_1-1)  t}{2 (2 - d + \nu_1 + \nu_3) (1 - d + \nu_1 + \nu_3)} \,.
\label{eq:olb-diag-1r}
\end{eqnarray}

Since the difference equation in the index $\nu_1$ is of second order, it will reduce any value of this index down to either $\nu_1=0$ or $\nu_1=1$. One, therefore, needs two equations in the index $\nu_3$, each one corresponding to one of the two boundary values of $\nu_1$:
\begin{eqnarray}
I_{1,1,\nu_3-2,1} &=& \frac{(4 - 2 \nu_3 + 2 t - d t + 2 \nu_3 t)}{2 (-2 + d - \nu_3)}I_{1,1,\nu_3-1,1} + \nonumber\\
&& \frac{(2 - d + 2 \nu_3)  (\nu_3-1)   t}{ 2 (3 - d + \nu_3) (2 - d + \nu_3)}I_{1,1,\nu_3,1} \,,\label{eq:olb-diag-2}\\
I_{0,1,\nu_3-1,1} &=& -\frac{(2 - d + 2 \nu_3)  t}{2 (2 - d +\nu_3)}I_{0,1,\nu_3,1}\,.
\label{eq:olb-diag-3}
\end{eqnarray}

The equations for the other bottom group of sectors $\overline{(1, 0, 1, 0)}$ can be inferred by symmetry from eqs.~(\ref{eq:olb-diag-1},\ref{eq:olb-diag-1r},\ref{eq:olb-diag-2},\ref{eq:olb-diag-3}) above.

\end{document}